\documentclass[manyauthors,nocleardouble,COMPASS]{cernphprep} 
\RequirePackage{lineno}
\usepackage{cernunits,cernchemsym,heppennames2,graphicx,subfig,amssymb,color,amsmath}
\usepackage{wrapfig,rotating}
\usepackage{amssymb,amsmath,array,array}
\usepackage{placeins}
\usepackage{cite}

\usepackage[bottom]{footmisc}

\newcommand{\be}{\begin{equation}}
\newcommand{\ee}{\end{equation}}
\newcommand{\bq}{\begin{eqnarray}}
\newcommand{\eq}{\end{eqnarray}}

\newcommand{\D}{\mathrm{d}}

\def\Vec#1{\mathpalette{\VVec}{#1}}                  
\def\VVec#1#2{\mbox{\boldmath$#1#2$\unboldmath}}

\begin{document}
\begin{titlepage}
\PHnumber{2018--xxx}
\PHdate{\today}
\DEFCOL{CDS-Library}
\title{Measurement of $P_T$-weighted Sivers asymmetries in 
leptoproduction of hadrons}
\Collaboration{The COMPASS Collaboration}
\ShortAuthor{The COMPASS Collaboration}

%



\begin{abstract}
The transverse spin asymmetries measured in semi-inclusive
leptoproduction of hadrons, when weighted with the hadron transverse momentum
 $P_T$, allow for the extraction of important transverse-momentum-dependent
distribution functions.  In particular, the weighted Sivers asymmetries provide
direct information on the Sivers function, which is a leading-twist distribution
that arises from a correlation between the transverse momentum of an unpolarised
quark in a transversely polarised nucleon and the spin of the nucleon.  Using
the high-statistics data collected by the COMPASS Collaboration in 2010 with a
transversely polarised proton target, we have evaluated two types of 
$P_T$-weighted Sivers asymmetries, which are both proportional to the product 
of the first
transverse moment of the Sivers function and of the fragmentation function. The
results are compared to the standard unweighted Sivers asymmetries and used to
extract the first transverse moments of the Sivers distributions for $u$ and $d$
quarks.
\end{abstract}



\vfill
\Submitted{(to be submitted to Nuclear Physics B)}
\end{titlepage}

{\pagestyle{empty}  
%
%
\section*{The COMPASS Collaboration}
\label{app:collab}
\renewcommand\labelenumi{\textsuperscript{\theenumi}~}
\renewcommand\theenumi{\arabic{enumi}}
\begin{flushleft}
M.G.~Alexeev\Irefn{turin_u},
G.D.~Alexeev\Irefn{dubna}, 
A.~Amoroso\Irefnn{turin_u}{turin_i},
V.~Andrieux\Irefnn{illinois}{saclay},
N.V.~Anfimov\Irefn{dubna}, 
V.~Anosov\Irefn{dubna}, 
A.~Antoshkin\Irefn{dubna}, 
K.~Augsten\Irefnn{dubna}{praguectu}, 
W.~Augustyniak\Irefn{warsaw},
C.D.R.~Azevedo\Irefn{aveiro}
% ...
\end{flushleft}
%
%
\begin{Authlist}
\item \Idef{alessandria}{Universita' degli Studi del Piemonte Orientale ``A. Avogadro'', Di.S.I.T., 15121 Alessandria, Italy}
\item \Idef{aveiro}{University of Aveiro, Dept.\ of Physics, 3810-193 Aveiro, Portugal}
\item \Idef{bochum}{Universit\"at Bochum, Institut f\"ur Experimentalphysik, 44780 Bochum, Germany\Arefs{l}\Aref{s}}
\item \Idef{bonniskp}{Universit\"at Bonn, Helmholtz-Institut f\"ur  Strahlen- und Kernphysik, 53115 Bonn, Germany\Arefs{l}}
\item \Idef{bonnpi}{Universit\"at Bonn, Physikalisches Institut, 53115 Bonn, Germany\Arefs{l}}
\item \Idef{brno}{Institute of Scientific Instruments, AS CR, 61264 Brno, Czech Republic\Arefs{m}}
\item \Idef{calcutta}{Matrivani Institute of Experimental Research \& Education, Calcutta-700 030, India\Arefs{n}}
\item \Idef{dubna}{Joint Institute for Nuclear Research, 141980 Dubna, Moscow region, Russia\Arefs{o}}
\item \Idef{freiburg}{Universit\"at Freiburg, Physikalisches Institut, 79104 Freiburg, Germany\Arefs{l}\Aref{s}}
\item \Idef{cern}{CERN, 1211 Geneva 23, Switzerland}
\item \Idef{liberec}{Technical University in Liberec, 46117 Liberec, Czech Republic\Arefs{m}}
\item \Idef{lisbon}{LIP, 1000-149 Lisbon, Portugal\Arefs{p}}
\item \Idef{mainz}{Universit\"at Mainz, Institut f\"ur Kernphysik, 55099 Mainz, Germany\Arefs{l}}
\item \Idef{miyazaki}{University of Miyazaki, Miyazaki 889-2192, Japan\Arefs{q}}
\item \Idef{moscowlpi}{Lebedev Physical Institute, 119991 Moscow, Russia}
\item \Idef{munichtu}{Technische Universit\"at M\"unchen, Physik Dept., 85748 Garching, Germany\Arefs{l}\Aref{r}}
\item \Idef{nagoya}{Nagoya University, 464 Nagoya, Japan\Arefs{q}}
\item \Idef{praguecu}{Charles University in Prague, Faculty of Mathematics and Physics, 18000 Prague, Czech Republic\Arefs{m}}
\item \Idef{praguectu}{Czech Technical University in Prague, 16636 Prague, Czech Republic\Arefs{m}}
\item \Idef{protvino}{State Scientific Center Institute for High Energy Physics of National Research Center `Kurchatov Institute', 142281 Protvino, Russia}
\item \Idef{saclay}{IRFU, CEA, Universit\'e Paris-Saclay, 91191 Gif-sur-Yvette, France\Arefs{s}}
\item \Idef{taipei}{Academia Sinica, Institute of Physics, Taipei 11529, Taiwan\Arefs{tw}}
\item \Idef{telaviv}{Tel Aviv University, School of Physics and Astronomy, 69978 Tel Aviv, Israel\Arefs{t}}
\item \Idef{triest_u}{University of Trieste, Dept.\ of Physics, 34127 Trieste, Italy}
\item \Idef{triest_i}{Trieste Section of INFN, 34127 Trieste, Italy}
\item \Idef{turin_u}{University of Turin, Dept.\ of Physics, 10125 Turin, Italy}
\item \Idef{turin_i}{Torino Section of INFN, 10125 Turin, Italy}
\item \Idef{tomsk}{Tomsk Polytechnic University,634050 Tomsk, Russia\Arefs{nauka}}
\item \Idef{illinois}{University of Illinois at Urbana-Champaign, Dept.\ of Physics, Urbana, IL 61801-3080, USA\Arefs{nsf}}
\item \Idef{warsaw}{National Centre for Nuclear Research, 00-681 Warsaw, Poland\Arefs{u}}
\item \Idef{warsawu}{University of Warsaw, Faculty of Physics, 02-093 Warsaw, Poland\Arefs{u}}
\item \Idef{warsawtu}{Warsaw University of Technology, Institute of Radioelectronics, 00-665 Warsaw, Poland\Arefs{u} }
\item \Idef{yamagata}{Yamagata University, Yamagata 992-8510, Japan\Arefs{q} }
\end{Authlist}
%
%
\renewcommand\theenumi{\alph{enumi}}
\begin{Authlist}
\item [{\makebox[2mm][l]{\textsuperscript{\#}}}] Corresponding authors
\item [{\makebox[2mm][l]{\textsuperscript{*}}}] Deceased
\item \Adef{a}{Also at Instituto Superior T\'ecnico, Universidade de Lisboa, Lisbon, Portugal}
\item \Adef{b}{Also at Dept.\ of Physics, Pusan National University, Busan 609-735, Republic of Korea and at Physics Dept., Brookhaven National Laboratory, Upton, NY 11973, USA}
\item \Adef{ictp}{Also at Abdus Salam ICTP, 34151 Trieste, Italy}
\item \Adef{r}{Supported by the DFG cluster of excellence `Origin and Structure of the Universe' (www.universe-cluster.de) (Germany)}
\item \Adef{p2i}{Supported by the Laboratoire d'excellence P2IO (France)}
\item \Adef{fuchey_adr}{Present address:  University of Connecticut, Storrs, Connecticut 06269, US}
\item \Adef{joerg_adr}{Present address: Universit\"at Bonn, Physikalisches Institut, 53115 Bonn, Germany}
\item \Adef{d}{Also at Chubu University, Kasugai, Aichi 487-8501, Japan\Arefs{q}}
\item \Adef{x}{Also at Dept.\ of Physics, National Central University, 300 Jhongda Road, Jhongli 32001, Taiwan}
\item \Adef{e}{Also at KEK, 1-1 Oho, Tsukuba, Ibaraki 305-0801, Japan}
\item \Adef{g}{Also at Moscow Institute of Physics and Technology, Moscow Region, 141700, Russia}
\item \Adef{h}{Present address: RWTH Aachen University, III.\ Physikalisches Institut, 52056 Aachen, Germany}
\item \Adef{yerevan}{Also at Yerevan Physics Institute, Alikhanian Br. Street, Yerevan, Armenia, 0036}
\item \Adef{y}{Also at Dept.\ of Physics, National Kaohsiung Normal University, Kaohsiung County 824, Taiwan}
\item \Adef{regen}{Also at Institut f\"ur Theoretische Physik, Universit\"at T\"ubingen, 72076 T\"ubingen, Germany}
\item \Adef{turin_p}{Also at University of Eastern Piedmont, 15100 Alessandria, Italy}
\item \Adef{i}{Present address: Uppsala University, Box 516, 75120 Uppsala, Sweden}
\item \Adef{c}{    Supported by the DFG Research Training Group Programmes 1102 and 2044 (Germany)} 
%
%
\item \Adef{l}{    Supported by BMBF - Bundesministerium f\"ur Bildung und Forschung (Germany)}
\item \Adef{s}{    Supported by FP7, HadronPhysics3, Grant 283286 (European Union)}
\item \Adef{m}{    Supported by MEYS, Grant LG13031 (Czech Republic)}
\item \Adef{n}{    Supported by B.Sen fund (India)}
\item \Adef{o}{    Supported by CERN-RFBR Grant 12-02-91500}
\item \Adef{p}{\raggedright 
                   Supported by FCT - Funda\c{c}\~{a}o para a Ci\^{e}ncia e Tecnologia, COMPETE and QREN, Grants CERN/FP 116376/2010, 123600/2011 
                   and CERN/FIS-NUC/0017/2015 (Portugal)}
\item \Adef{q}{    Supported by MEXT and JSPS, Grants 18002006, 20540299, 18540281 and 26247032, the Daiko and Yamada Foundations (Japan)}
\item \Adef{tw}{   Supported by the Ministry of Science and Technology (Taiwan)}
\item \Adef{t}{    Supported by the Israel Academy of Sciences and Humanities (Israel)}
\item \Adef{nauka}{Supported by the Russian Federation  program ``Nauka'' (Contract No. 0.1764.GZB.2017) (Russia)}
\item \Adef{nsf}{  Supported by the National Science Foundation, Grant no. PHY-1506416 (USA)}
\item \Adef{u}{    Supported by NCN, Grant 2017/26/M/ST2/00498 (Poland)}
\end{Authlist}

\clearpage
}

\section{Introduction}
The traditional description of the nucleon structure in hard inclusive processes
in terms of collinear parton distributions functions, which depend on the parton
light-cone momentum fraction $x$ and on a characteristic hard scale $Q^2$, was
recently generalised to take into account the transverse momentum $k_T$ of the
parton with respect to the nucleon direction (for reviews, 
see~\cite{Barone:2010zz,Aidala:2012mv,Avakian:2016rst}).  
A complete picture of the
nucleon at leading twist requires a total of eight transverse-momentum-dependent
distributions (TMDs).  
They provide important information on the dynamics of the
partons in the transverse plane in momentum space. 
Upon integration over the
transverse momentum, three of them reduce to the number density, the helicity
and the transversity collinear distributions.  The other five TMDs contain
prefactors that are sensitive to the direction of the quark transverse momentum
vector $\Vec{k}_T$, and their contribution to the hadronic tensor vanishes when
integrating over $\Vec{k}_T$.

Among the TMDs, an important r{\^o}le is played by the Sivers distribution function
$f_{1T}^{\perp \, q}$~\cite{Sivers:1989cc,Sivers:1990fh,Brodsky:2002rv,Collins:2002kn},
which for an
unpolarised quark of flavour $q$ describes the correlation between 
its transverse momentum and the transverse polarisation of the nucleon.  
In semi-inclusive measurements of deep-inelastic scattering (SIDIS) 
off a transversely polarised
nucleon, the Sivers TMD embodies in the cross section a sine modulation on the
difference between the azimuthal angle $\phi_h$
of the produced hadron and that of
the target nucleon spin, $\phi_S$.

The Sivers effect was experimentally observed in SIDIS using transversely
polarised proton targets, first by the HERMES 
Collaboration~\cite{Airapetian:2004tw,Airapetian:2009ae} and then, at higher energy, by the
COMPASS Collaboration~\cite{Alekseev:2010rw,Adolph:2012sp}. 
The COMPASS measurements on the deuteron~\cite{Alexakhin:2005iw,Ageev:2006da} showed asymmetries
compatible with zero within the experimental accuracy.  More recently, data on
pion production off a transversely polarised $^3$He target were made available
by the Hall A Collaboration at JLab~\cite{Qian:2011py}.  Combined analyses of
these measurements~\cite{Efremov:2003tf,Efremov:2004tp,Collins:2005ie,Vogelsang:2005cs,Anselmino:2005nn,Anselmino:2005ea,Anselmino:2005kn,Anselmino:2008sga,Anselmino:2012aa,Sun:2013dya,Echevarria:2014xaa}
allowed for extractions of the Sivers functions and of their first transverse moments
$f_{1T}^{\perp \, (1) \, q}$:
\begin{eqnarray}
f_{1T}^{\perp \, (1) \, q}(x) = \int d^2 \Vec{k}_T \, \frac{k_T^2}{2 M^2} \, \,
f_{1T}^{\perp \, q}(x, k_T^2) ,
\label{k_T_moment}
\end{eqnarray}
which are found to be different from zero, a very
important result in TMD physics. In particular, the $u$ and the $d$
distributions turn out to have similar magnitude, but opposite sign. In
Eq.~(\ref{k_T_moment}), $M$ is the target nucleon mass.  

While in most phenomenological studies the first transverse moments of the
Sivers distributions are extracted by fitting the data using a given functional
form for the $x$ dependence of $f_{1T}^{\perp}$, in Ref.~\cite{Martin:2017yms} a
different approach was adopted: the COMPASS measurements on proton and deuteron
targets in the same kinematics were used to extract point-by-point the first
transverse moments of the Sivers distributions $f_{1T}^{\perp (1)}$ directly
from the data by combining the various asymmetries.

The main problem in all extractions performed up to 
now is that the standard Sivers
asymmetries involve transverse-momentum convolutions of TMDs and fragmentation
functions, from which the first transverse moments of the Sivers functions can
be obtained analytically only by assuming a specific form, typically a Gaussian,
for the transverse-momentum dependence of all involved quantities.

Already twenty years ago an alternative method was 
proposed~\cite{Kotzinian:1995cz,Kotzinian:1997wt,Boer:1997nt} to determine 
$f_{1T}^{\perp  (1)}$ without making any assumption on the functional form of
the transverse-momentum 
dependence, neither for the distribution functions nor for the
fragmentation functions.  
The method, which consists in measuring asymmetries
weighted by the measurable transverse momentum $P_{T}$ of the hadron, was not
pursued; the only and still preliminary results came from
HERMES~\cite{Gregor:2005qv}.  
It is worth to mention that the first transverse moment of the Sivers function
is directly entering in the Burkardt sum rule~\cite{Burkardt:2004ur}, which
allows to constrain the gluon Sivers function using the  measured
Sivers functions for quarks~\cite{Anselmino:2008sga}.  
Recently, much interest has been
dedicated again to the weighted asymmetries (see
e.g.~\cite{Kang:2012ns,Hadjidakis:2018ifr}). 

In this paper, we
present the first measurements of two types of $P_T$-weighted Sivers asymmetries
performed by the COMPASS collaboration using the high statistics data collected
in 2010 with a  160 GeV muon beam impinging on a transversely polarised proton 
target. The
results are compared to the standard unweighted Sivers asymmetries and used to
extract the first transverse moments of the Sivers functions for $u$ and $d$
quarks.

\section{The Sivers asymmetries}
The Sivers asymmetry is associated to a $\sin \Phi_{\rm Siv} \equiv \sin (\phi_h
- \phi_S)$ modulation of the SIDIS cross section in a reference frame where the
momentum vectors of virtual photon and nucleon are collinear, 
the $z$ axis is taken along the
virtual-photon momentum and the $x$ axis  along the lepton transverse
momentum.  The relevant part of the fully differential cross section is
\be
\D \sigma = \D \sigma_U + S_T \, \D \sigma_S \, \sin \Phi_{\rm Siv},
\ee
where $S_T$ is the target nucleon polarisation, and $\D \sigma_U$ and $\D
\sigma_S$ are the spin-independent and spin-dependent parts of the 
cross section, respectively.
In the standard, i.e. unweighted case, the Sivers asymmetry is defined as
\be
A_{\rm Siv} = 2 \, \frac{\int \D \Phi_{\rm Siv} \D \phi_h \, \sin \Phi_{\rm Siv}
\, \D \sigma}{\int \D \Phi_{\rm Siv} \D \phi_h
\, \D \sigma}\, .
\label{standard_asymmetry}
\ee
At leading twist and leading order in QCD, $A_{\rm Siv}$ is 
given~\cite{Mulders:1995dh,Boer:1997nt} 
in terms of the Sivers function $f_{1T}^{\perp}$ and the 
transverse-momentum-dependent unpolarised distribution
and fragmentation functions $f_1$ and $D_1$ by
\be
 A_{\rm Siv}(x,z,P_T)
 = 
\frac{\sum_{q} e_q^2 x 
\, \mathcal{C} \left [ 
\frac{\Vec P_T \cdot \Vec k_T}{M P_T} \,   
f_{1T}^{\perp  \, q}(x, k_T^2) \, D_1^q(z, p_T^2) \right ] }
     {\sum_{q} e_q^2 x 
\, \mathcal{C} \left [  f_{1}^q(x, k_T^2) D_1^q(z, p_T^2) \right ]} ,  
\label{eq:sa}
\ee
where the sums are over quark and antiquark flavours, $e_q$ are the 
quark charges, and the
transverse momentum convolutions are given by
\bq
& & \mathcal{C}  \left [  \frac{\Vec P_T \cdot \Vec k_T}{M P_T} \,   
f_{1T}^{\perp  \, q} \, D_1^q  \right ] 
\nonumber \\
& & \;\;\;\;
\equiv 
\int \D^2 \Vec k_T  \int \D^2 \Vec p_T \, 
\delta^2 (z \Vec k_T + \Vec p_T - \Vec P_T) \, 
\frac{\Vec P_T \cdot \Vec k_T}{M P_T} \,   
f_{1T}^{\perp  \, q}(x, k_T^2) \, D_1^q(z, p_T^2)\,, 
\label{conv_num}
\eq
and
 \be
 \mathcal{C}  \left [   
f_{1}^{ q} \, D_1^q  \right ] 
\equiv 
 \int \D^2 \Vec k_T  \int \D^2 \Vec p_T \, 
\delta^2 (z \Vec k_T + \Vec p_T - \Vec P_T) \,    
f_{1}^{q}(x, k_T^2) \, D_1^q(z, p_T^2)\,.  
\label{conv_den}
\ee
In Eqs.~(\ref{eq:sa},\ref{conv_num},\ref{conv_den}), $z$ is the fraction of the
longitudinal momentum of the fragmenting quark carried by the produced hadron,
$\Vec p_T$ is the transverse momentum of the produced hadron with respect to the
direction of the fragmenting quark momentum.
For simplicity, we have omitted the $Q^2$ dependence of  parton
distributions, fragmentation functions and Sivers asymmetry.

When integrating over $\Vec P_T$, the denominator of Eq.~(\ref{eq:sa}) is easily
computed yielding the familiar ``collinear'' expression
\be
\sum_{q} e_q^2 x \, \int \D^2 \Vec P_T 
\, \mathcal{C} \left [  f_{1}^q D_1^q \right ]
= 
\sum_{q} e_q^2 x \, f_{1}^q(x) D_1^q(z)\,, 
\ee
where $f_1^q (x)$ and $D_1^q(z)$ are the above defined partonic functions
integrated over the transverse momentum, while, in the general case, the numerator of
Eq.~(\ref{eq:sa}) cannot be analytically evaluated.  
Hence, in order to disentangle
$f_{1T}^{\perp}$ and $D_1$ and to extract the Sivers function, some functional
form must be assumed for the transverse-momentum dependence of the distribution
and fragmentation functions. Assuming this form to be a Gaussian, the Sivers
asymmetry becomes~\cite{Boer:1997nt,Efremov:2003tf,Efremov:2004tp}
\begin{eqnarray}
A_{{\rm Siv},{\rm G}}(x,z) &=& 
\frac{a_{\rm G} \, \sum_{q} e_q^2 x  f_{1T}^{\perp \, (1) \, q}(x) z  D_1^q(z)}
     {\sum_{q} e_q^2 x  f_{1}^q(x) D_1^q(z)} \,.
\label{eq:gsa}
\end{eqnarray}
The factor $a_G$ in Eq.~(\ref{eq:gsa}) is  
\be
a_{\rm G} = 
\frac{\sqrt{\pi} M}{\sqrt{\langle p_T^2 \rangle + 
z^2 \langle k_T^2 \rangle_S}}\,,  
\label{eq:aG}  
\ee
where $\langle p_T^2 \rangle$ and $\langle k_T^2 \rangle_S$ are the Gaussian
widths of the fragmentation function and of the Sivers function, respectively.
In the Gaussian model, the average transverse momentum of the produced hadrons
(integrated over its azimuthal angle) is written as
\be
\langle P_{T} \rangle = \frac{\sqrt{\pi}}{2} 
\sqrt{\langle p_T^2 \rangle + z^2 \langle k_T^2 \rangle}\,,   
\label{PT}
\ee
where $\langle k_T^2 \rangle$ is the width of the transverse-momentum-dependent
number density $f_1$, which in principle differs from $\langle k_T^2 \rangle_S$.
Taking approximately $\langle k_T^2 \rangle_S \simeq \langle k_T^2 \rangle$, we
can write $a_{\rm G}$ as
\be
a_{\rm G} \simeq \frac{\pi M}{2 \langle P_{T} \rangle}. 
\label{eq:aG2}
\ee
The Gaussian ansatz clearly introduces a bias into the extraction of the Sivers
function.  In order to avoid this problem one can consider, instead of
Eq.~(\ref{standard_asymmetry}), an asymmetry that is weighted by the transverse
momentum of the produced hadron.  In particular, when choosing $w = P_T/zM$ as
weight, the weighted Sivers asymmetry becomes
\be
A_{\rm Siv}^w = \frac{\int \D \Phi_{\rm Siv} \, \sin \Phi_{\rm Siv}\,
\int \D^2 \Vec P_T \, \left (\frac{P_T}{z M} \right ) \,  \D \sigma}{\int \D \Phi_{\rm Siv} 
 \int \D^2 \Vec P_T 
\, \D \sigma}\,.     
\label{weighted_asymmetry}
\ee
In terms of quark distribution and fragmentation functions, it reads
\be
 A_{\rm Siv}^w(x,z) 
 = 
\frac{\sum_{q} e_q^2 x \, \int \D^2 \Vec P_T 
\, \frac{P_T}{z M} \, \mathcal{C} \left [ 
\frac{\Vec P_T \cdot \Vec k_T}{M P_T} \,   
f_{1T}^{\perp  \, q}(x, k_T^2) \, D_1^q(z, p_T^2) \right ] }
     {\sum_{q} e_q^2 x \,  f_{1}^q(x) D_1^q(z) } ,  
\label{eq:W_sa}
\ee
The convolution in the numerator can now be carried out in a straightforward way
(see Appendix A) and the final expression is
\be
A_{\rm Siv}^w (x,z) = 2 \, 
\frac{\sum_{q} e_q^2 x  f_{1T}^{\perp \, (1) \, q}(x)   D_1^q(z)}
     {\sum_{q} e_q^2 x  f_{1}^q(x) D_1^q(z)} \,, 
\label{eq:w_sa_esplicit}
\ee
which shows that the asymmetry contains the product of the first $k_T^2$ moment
of the Sivers function and the unpolarised fragmentation function.

When using $w' = P_T/M$ as weight, the resulting Sivers asymmetry reads
\be
A_{\rm Siv}^{w'} = \frac{\int \D \Phi_{\rm Siv} \, \sin \Phi_{\rm Siv}\,
\int \D^2 \Vec P_T \, \left (\frac{P_T}{M} \right ) \,  \D \sigma}{\int \D \Phi_{\rm Siv} 
 \int \D^2 \Vec P_T 
\, \D \sigma}\,.     
\label{weighted_asymmetry_bis}
\ee
This asymmetry is of interest because it should exhibit a $z$ dependence close
to that of the unweighted asymmetries. Its expression in the parton model,
\begin{eqnarray}
A_{\rm Siv}^{w'}(x,z) &=& 
2 \frac{\sum_{q} e_q^2 x f_{1T}^{\perp \, (1) \, q}(x) \, z D_1^q(z)}
     {\sum_{q} e_q^2 x f_{1}^q(x) \, D_1^q(z)}\,, 
\label{eq:wpsa}
\end{eqnarray}
is indeed very similar to that of the unweighted asymmetry in the Gaussian
model, Eq.~(\ref{eq:gsa}).  In particular, from
Eqs.~(\ref{eq:gsa},\ref{eq:aG2},\ref{eq:wpsa}) one sees that the ratio $
A_{\rm Siv}^{w'}/A_{\rm Siv,G}$ is related to the average value of the hadron transverse
momentum:
\be
\frac{A_{\rm Siv}^{w'}}{A_{\rm Siv, G}} 
 \simeq \frac{4 \langle  P_T \rangle}{\pi M}\,. 
\label{ratio}
\ee

\section{Experimental set-up and data analysis}
The COMPASS spectrometer~\cite{Abbon:2007pq,Abbon:2014aex} 
is in operation in the SPS North
Area of CERN since 2002. The data used in this analysis were collected in 2010
by scattering a 160 GeV $\mu^+$ beam on a transversely polarised 
target. The 1.2~m long ${\rm NH}_3$ target was kept at 50~mK in a dilution
refrigerator cryostat and segmented in three cells, 30~cm, 60~cm and 30~cm long
respectively.  The proton polarisation of about 80\%
was oriented vertically by a 0.63 T
magnetic field that was provided by the saddle coils of the polarised target 
magnet~\cite{Koivuniemi:2015uyw}.
The data were taken at a mean beam intensity of $3.5\times
10^8$ $\mu$/spill, for a spill length of about 10 s every 40 s.  About $37\times
10^9$ events, corresponding to 1.9 PB of data, were collected in twelve separate
periods.  
In order to minimize systematic errors, during each period of data
taking the orientation of the proton polarisation in the three target
cells was either up-down-up or down-up-down in the first subperiod, and
reversed in the second one.  By suitably combining the data, instrumental
asymmetries could be limited to negligible values.  The principles of the
measurement and the data analysis were already described in several 
publications~\cite{Alekseev:2010rw,Adolph:2012sp,Alexakhin:2005iw} and will not be repeated
here.

In order to allow for a comparison of the weighted Sivers asymmetries with the
unweighted asymmetries, all constraints to select DIS events and final-state hadrons
are the same as for the published data~\cite{Adolph:2012sp}.  Here we only recall
that in order to ensure the DIS regime only events with photon virtuality $Q^2 >1 \,
({\rm GeV}/c)^2$, fractional energy of the virtual photon $ 0.1 < y < 0.9$, and
mass of the hadronic final-state system $ W > 5$ GeV/$c^2$ are considered.  A
charged hadron is required to have a transverse momentum $P_T \ge 0.1$ GeV/$c$
and a fraction of the available energy $z>0.2$. With these constraints, about $8\times
10^7$ hadrons are left and used for the extraction of the asymmetries.  This
sample consists mainly
of pions (about 70\% for positive hadrons, 75\% for negative 
hadrons~\cite{Adolph:2014zba}).  
In addition, the analysis was also done for charged
hadrons in the region $0.1<z<0.2$.

The weighted asymmetries are measured separately for positive and negative
hadrons as a function of $x$ or $z$.  For each bin in $x$ or $z$ and for each
period of data taking, the asymmetries are extracted from the number of hadrons
produced in each cell for the two directions of the target polarisation, and the
mean of the results from the twelve periods is taken as final result.

The unweighted asymmetries were extracted using both an extended unbinned
maximum likelihood method and the so-called double ratio method (DRM).  The two
methods led to very similar results and the small differences were added to the
systematic uncertainties.  In both cases, the hadrons produced in the two
data-taking subperiods and in the three target cells are combined in order to
ensure cancellation of the azimuthal acceptance and of the beam flux.  Since
only the counts in the numerator of the expression of $A_{\rm Siv}^{w}$ are
weighted, a modified DRM is used in this analysis.

In each kinematic bin, we divide the $\Phi_{\rm Siv}$ range in 12 bins, 
and in each of them we calculate the quantity
\begin{eqnarray}
R(\Phi_{\rm Siv}) &=&
\frac{\Delta^w}{\sqrt{\Sigma^w \Sigma}} \, ,
\label{eq:rratio}
\end{eqnarray}
where
\begin{equation}
\Delta^w = N_+^w N_+^{'w} - N_-^{'w} N_-^w  \, , \; \; \; 
\Sigma^w = N_+^w N_+^{'w} + N_-^{'w} N_-^w  \, , \; \; \; 
\Sigma = N_+ N'_+ + N'_- N_- \, .
\end{equation}
Here $N$ and $N^w$ are the number of counts and the sum of weights, 
respectively,  and $N$ ($N'$) refers to the first (second)
subperiod.  
The numbers of hadrons produced in the first and in the third target cell,
which are always polarised in the same direction, are added up.  
The subscripts $+$ and $-$ indicate the up and down 
orientation of the target polarisation. 
Both azimuthal
acceptance and beam flux cancel in the ratio of Eq.~(\ref{eq:rratio}), so that
\begin{eqnarray}
R(\Phi_{\rm Siv}) &\simeq& 
4 \bar{S}_T \, A^{w}_{\rm Siv} \, \sin \Phi_{\rm Siv} ,
\end{eqnarray}
where $\bar{S}_T$ is the mean transverse polarisation of the target protons.
Cancellation of azimuthal acceptance is guaranteed as long as the ratios of the
acceptances of the oppositely polarised cells in the two data taking subperiod
are the same, which is the so-called ``reasonable 
assumption''~\cite{Ageev:2006da}.

Several tests were performed to assess the correctness of the results and the
size of possible systematic uncertainties.
Two alternative estimators were used, which are not expected to guarantee an as
good cancellation of the azimuthal acceptance as the modified DRM but are much
simpler, one of them being the mean value of $\sin \Phi_S \, P_T/zM$. It turned
out that the results are essentially identical.

The effect of the $P_T/z$ acceptance was also investigated.  
This acceptance is
about 60\% and rather flat in the range $0.020<x<0.7$ both for positive and
negative hadrons.  At smaller $x$ it increases smoothly from 0.4 to about 0.8 as
$P_T/z$ increases from 0.1~GeV$/c$ to 10~GeV$/c$.  
In order to evaluate the effect
of the acceptance in the results, we have re-evaluated $A^{w}_{\rm Siv}$ after
having corrected for the
$P_T/z$ acceptance.  The difference between the results obtained with
and without the corrections is at most one tenth of a standard deviation, and
thus negligible.

The stability of the results was checked paying particular attention to the
$P_T$ limits.  The effect of the lower $P_T$ cut, which is expected to be
negligible, was investigated by extracting the weighted Sivers asymmetries using three
different lower cuts, $P_T>0.15$~GeV/$c$, $P_T>0.20$ GeV/$c$ and
$P_T>0.25$~GeV/$c$.  Also, the effect of a cut on the upper value of $P_T$ was
investigated by extracting the asymmetries using the 
limits: $P_T<1.5$ GeV/$c$, $P_T<1.25$
GeV/$c$ and $P_T<1.0$ GeV/$c$.  
In all the cases the differences to the results
obtained with the standard cuts are negligibly small in all $x$ bins.

The contributions from higher-order processes, i.e.
QCD Compton and photon-gluon fusion, which are more relevant at high
$P_T$~\cite{Adolph:2015cvj}, have neither been taken into account
nor corrected for.

Altogether, no evidence was found for additional relevant systematic
uncertainties.  
The systematic uncertainties are estimated to be half of the
statistical uncertainties, 
as in the analysis of the standard Sivers asymmetries of the
same data~\cite{Adolph:2012sp}.

\section{Sivers asymmetries weighted by $P_T/zM$ }
\begin{figure*}[b]
\centering
\includegraphics[width=0.49\textwidth]{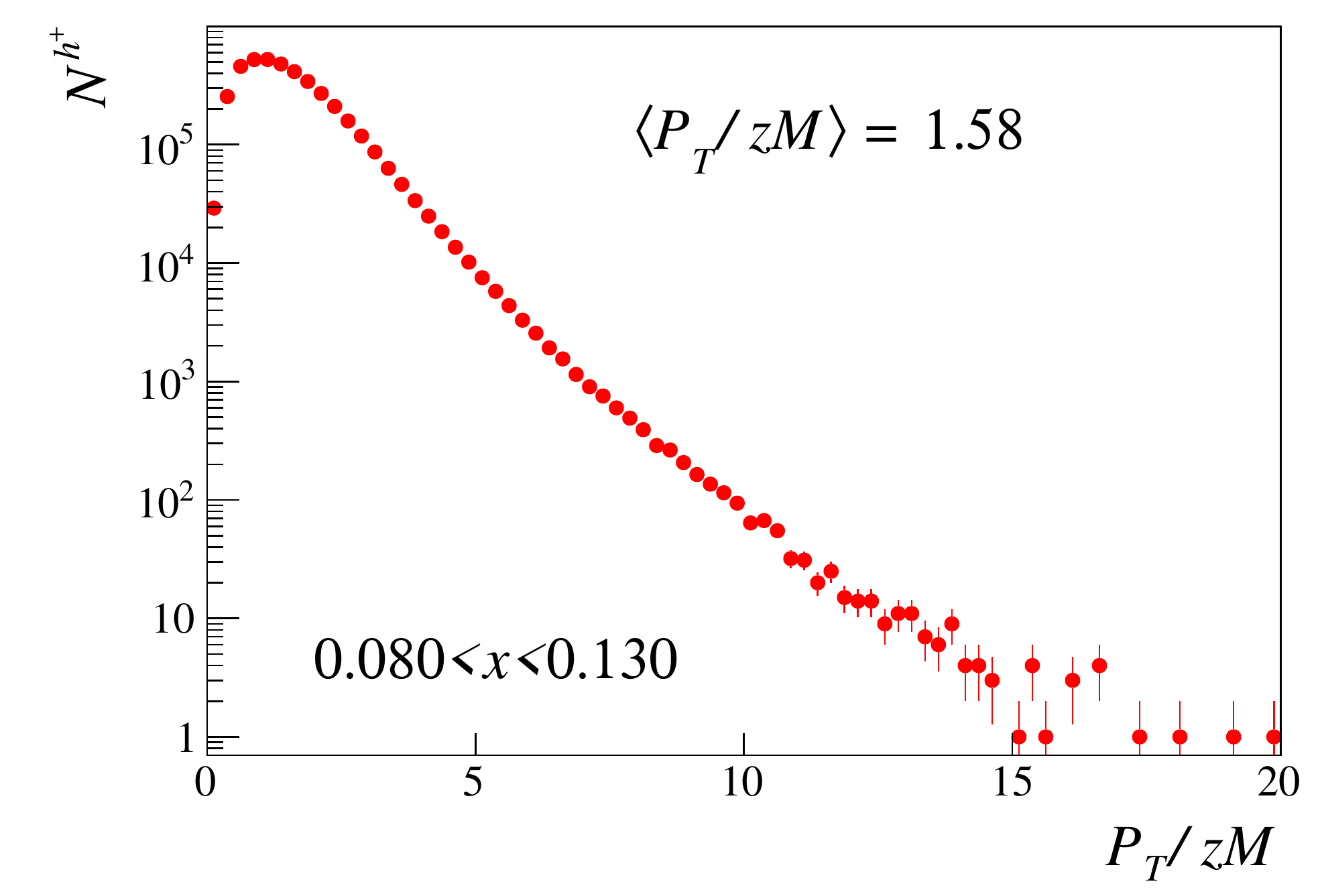}
\includegraphics[width=0.49\textwidth]{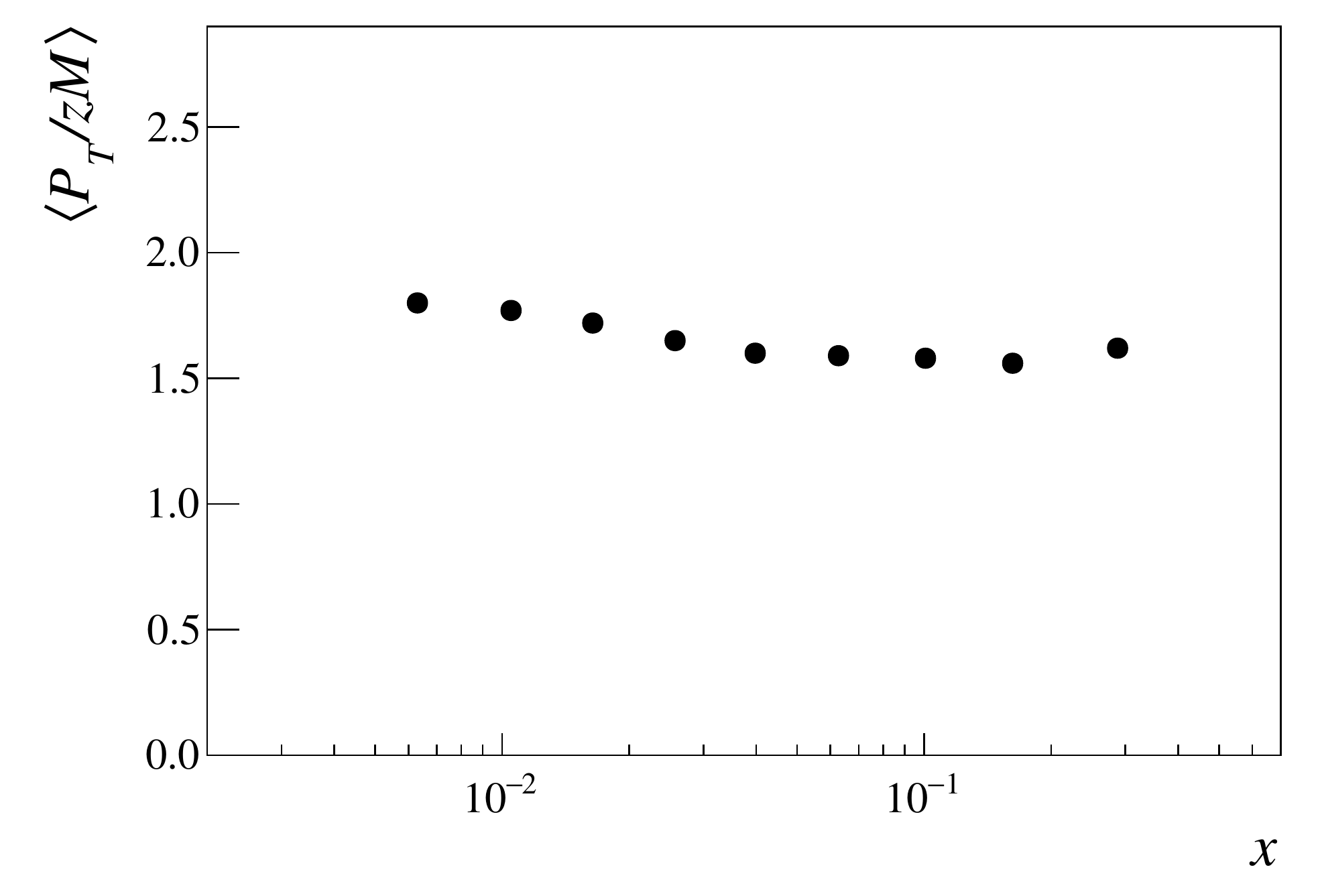}
\hfill
\caption{Left panel: Distribution of the weight  $w=P_T/zM$  for positive hadrons 
in the bin $0.080<x<0.13$. 
Right panel: Mean value of $w$ as function of $x$. No acceptance correction
applied.}
\label{fig:weightsptz_x}
\end{figure*}

\begin{figure*}[tb]
\centering
\includegraphics[width=0.48\textwidth]{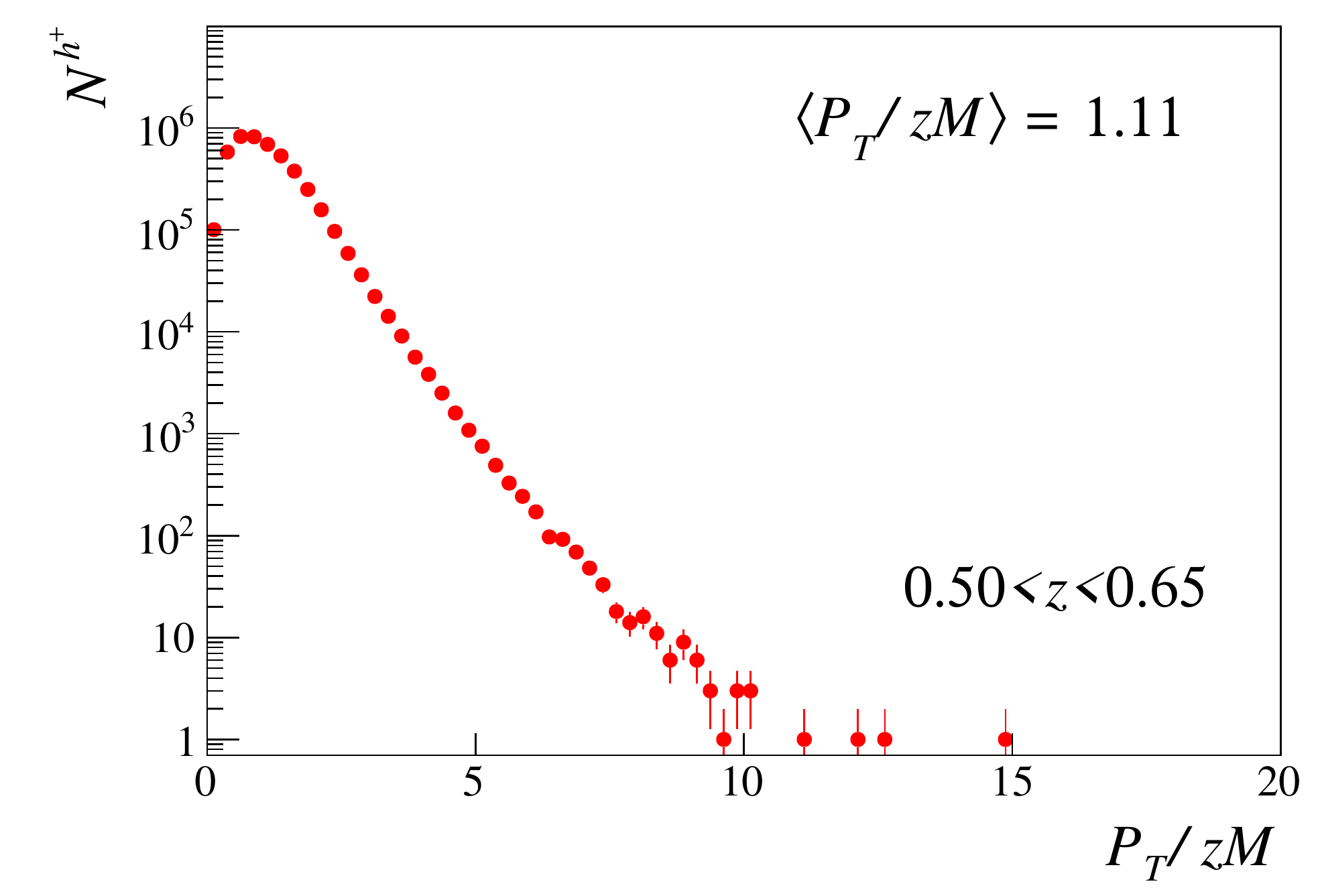}
\includegraphics[width=0.48\textwidth]{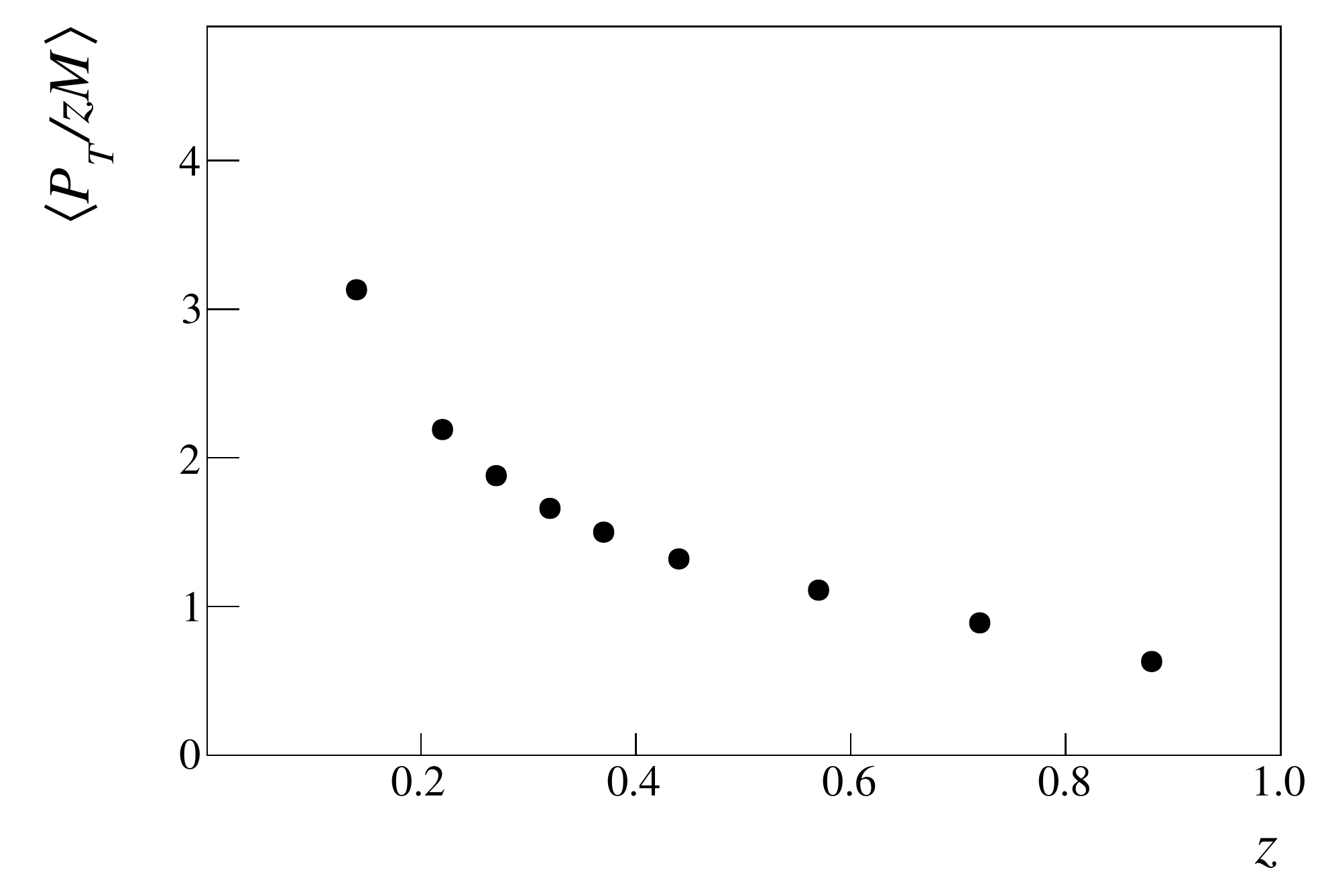}
\hfill
\caption{Left panel: Distribution of the weight  $w=P_T/zM$  for positive hadrons 
in the bin $0.50<z<0.65$. 
Right panel: Mean value of $w$ as function of $z$. 
No acceptance correction applied.}
\label{fig:weightsptz_z}
\end{figure*}

The distributions of the weights $w=P_T/zM$ are 
very similar for all nine $x$ bins.  
As an example, the distribution for positive hadrons
in the bin $0.080<x<0.130$ is
shown in the left panel of Fig.~\ref{fig:weightsptz_x}.  The mean values of $w$
in the nine $x$ bins are given in the right panel of the same
figure and in Table~\ref{tab:weightsptz}.  For negative hadrons the
distributions are very much the same.
The distributions of $w$ in the nine $z$ bins have also similar shapes 
but different slopes.  
The distribution for $0.50<z<0.65$ and the mean values of $w$ as
function of $z$ are shown in Fig.~\ref{fig:weightsptz_z} for positive hadrons.
Again, for negative hadrons the distributions are very much the same.

\begin{figure*}[tb]
\centering
\includegraphics[width=0.48\textwidth]{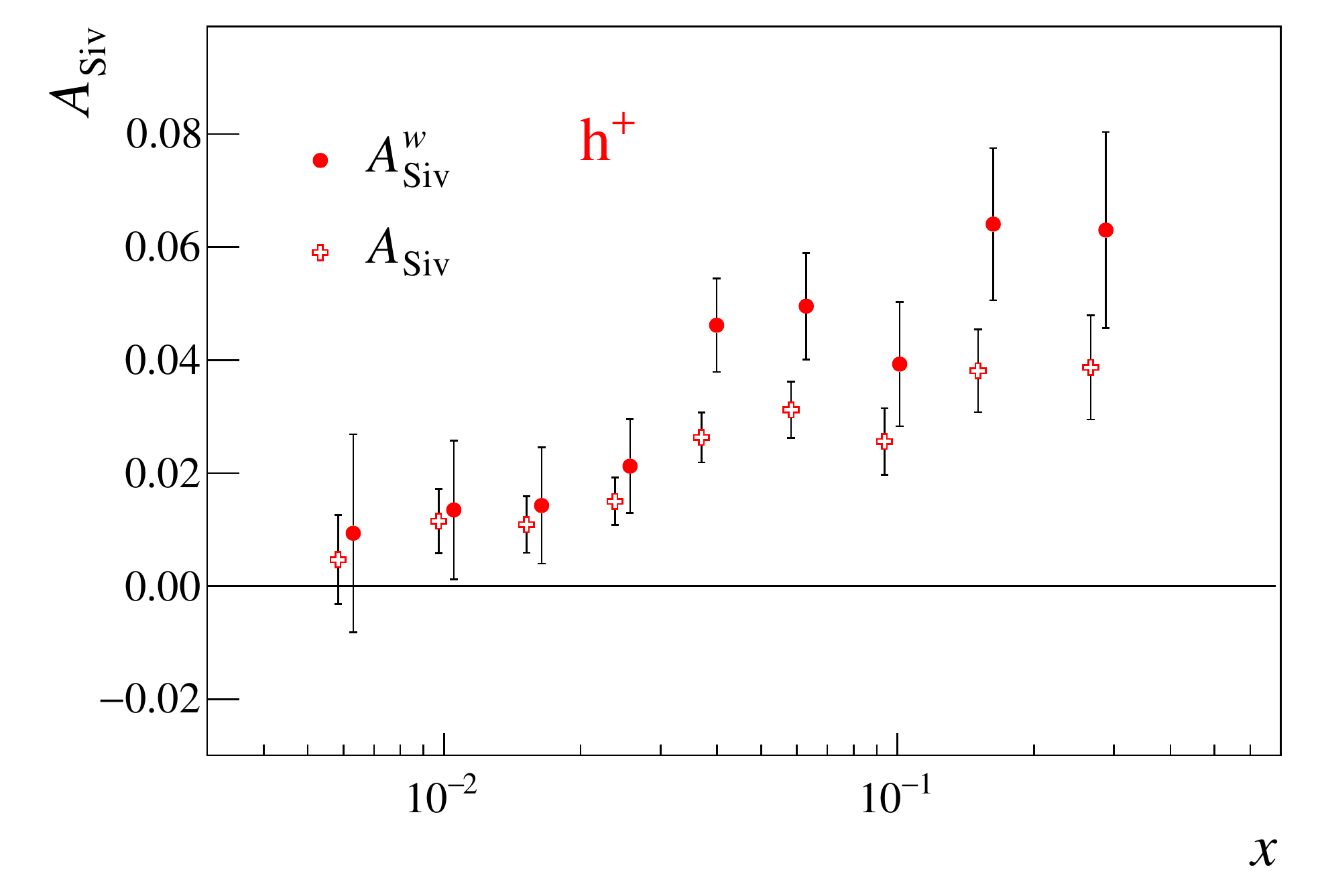}
\includegraphics[width=0.48\textwidth]{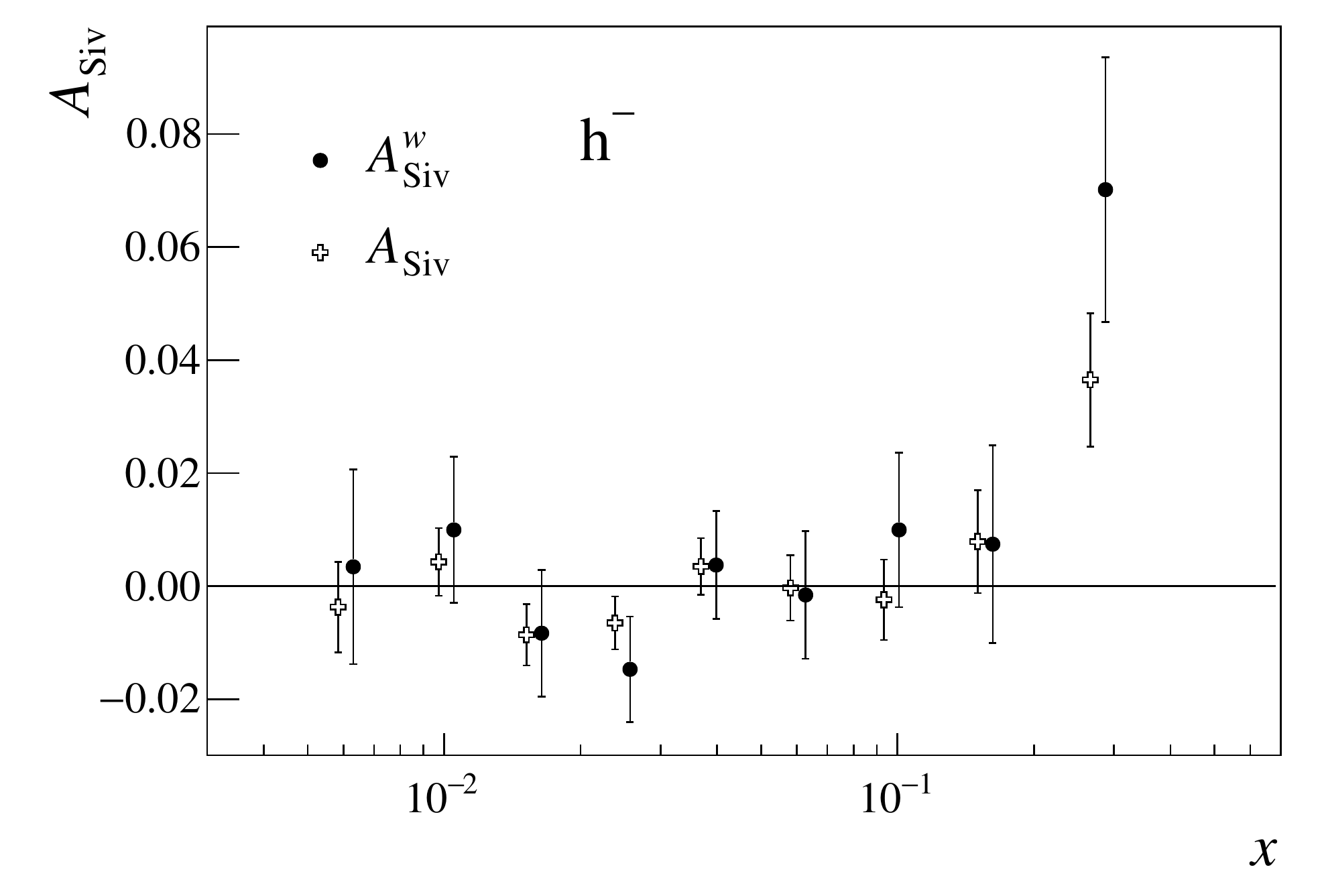}
\hfill
\caption{Full points: $A_{\rm Siv}^{w}$ in the nine $x$ bins for positive 
(left panel) and negative (right panel) hadrons. The open crosses are the
unweighted Sivers asymmetries  $A_{\rm Siv}$~\cite{Adolph:2012sp},
which are slightly shifted towards smaller $x$ values for clarity.}
\label{fig:results}
\end{figure*}

The measured weighted asymmetries are presented as a function of $x$ in
Fig.~\ref{fig:results}.
The unweighted Sivers asymmetries~\cite{Adolph:2012sp} are also shown for
comparison.  As expected, the trends of the weighted and unweighted asymmetries
are similar both for positive and negative hadrons.  The asymmetry for positive
hadrons is clearly different from zero, in particular at large $x$.  In this
range, the ratios $A_{\rm Siv}^w/A_{\rm Siv}$ are very close to the mean value
of the weight. The statistical uncertainties are scaled by about the same
ratio.

Assuming $u$-quark dominance for positive hadrons produced on a proton target,
one has
\be
A_{\rm Siv}^w \simeq 2 \, \frac{f_{1T}^{\perp (1) u}(x, Q^2)}{f_1^u (x, Q^2)}\,,  
\ee 
and  the results on $A_{\rm Siv}^w$ represent the first direct
measurement of $f_{1T}^{\perp \, (1) \, u}/ f_{1}^u$.

\begin{figure*}[tb]
\centering
\includegraphics[width=0.48\textwidth]{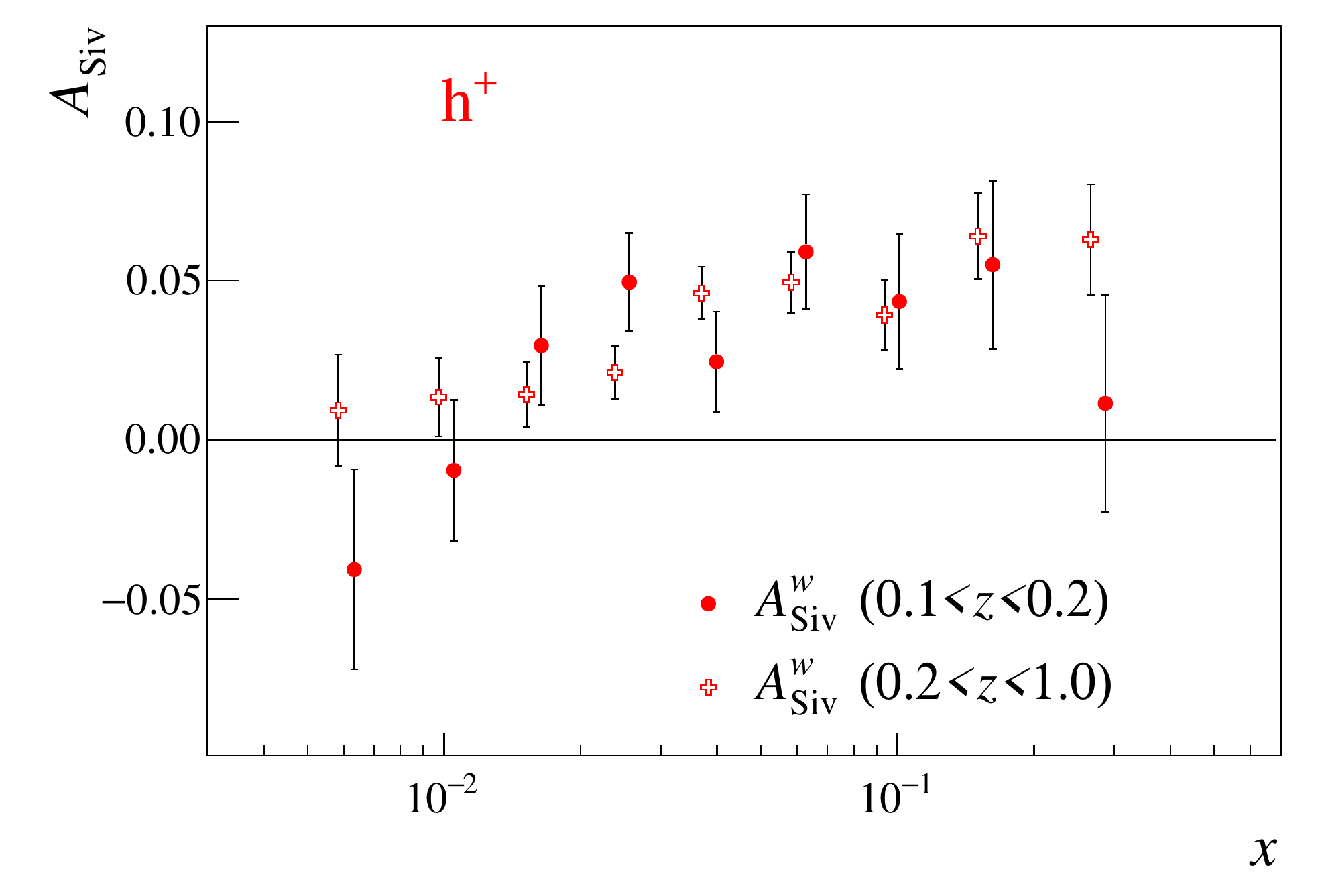}
\includegraphics[width=0.48\textwidth]{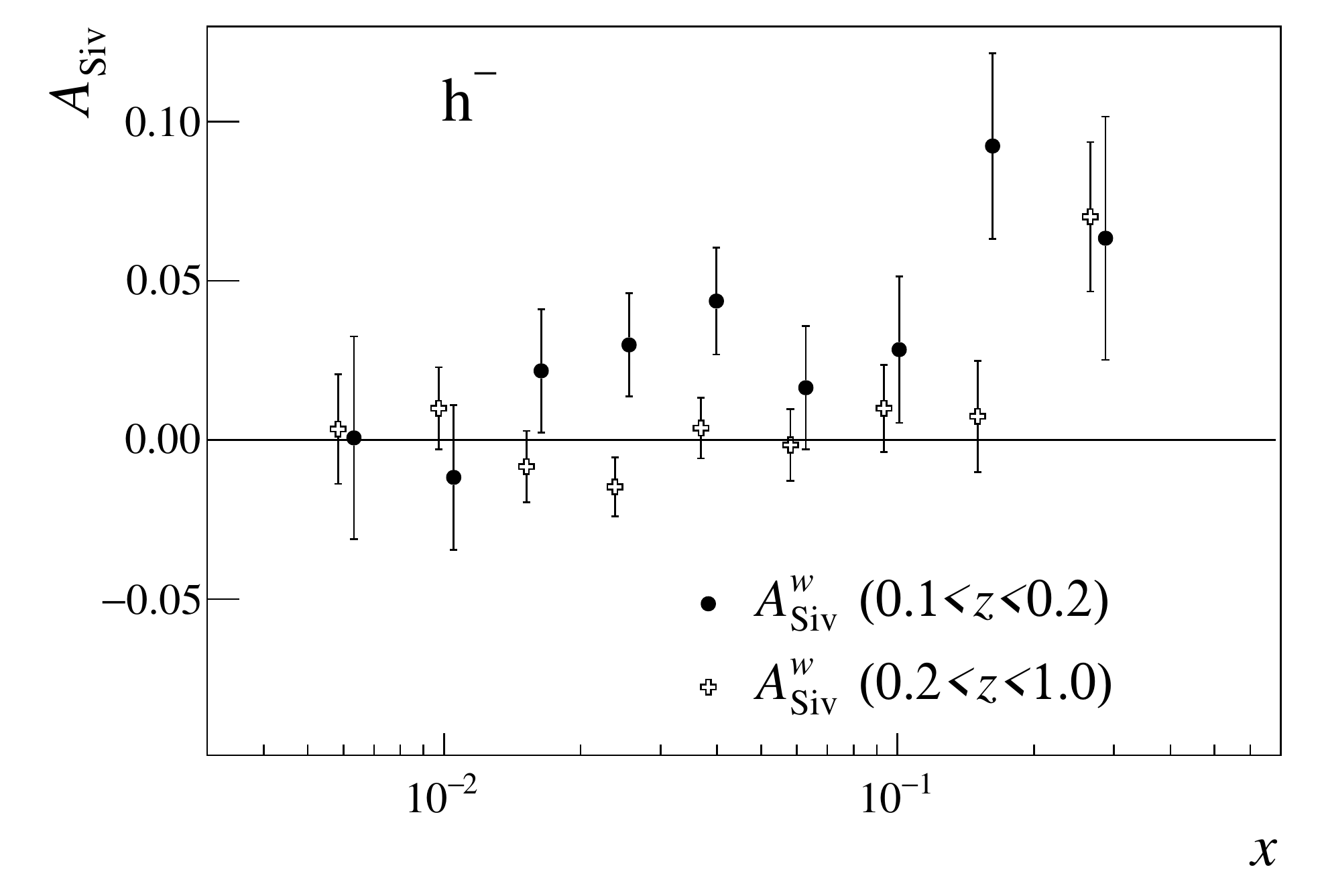}
\caption{Comparison of the weighted asymmetries 
vs. $x$ measured in the range 
($0.1<z<0.2$) for positive (left) and negative (right) hadrons
and the corresponding ones in the standard 
range $z>0.2$, which are slightly shifted towards smaller $x$ values for clarity.}
\label{fig:lzhz}
\end{figure*}

In Fig.~\ref{fig:lzhz}, the weighted Sivers asymmetries measured in our standard
range $z>0.2$ are compared with the corresponding ones in the range $0.1<z<0.2$.
It is interesting to note that the positive-hadron asymmetries are basically
unchanged, which emphasizes $u$-quark dominance and 
supports the idea that factorisation works already at small values of $z$ in the COMPASS kinematic
range.  At low $z$, the difference between favoured and unfavoured fragmentation
functions decreases, thus it is expected that the $u$-quark contribution to the
negative-hadron asymmetry increases.  The asymmetry itself is 
then expected to become
larger and similar to the positive-hadron asymmetries, as observed in
Fig.~\ref{fig:lzhz}.

\begin{figure*}[tb]
\centering
\includegraphics[width=0.48\textwidth]{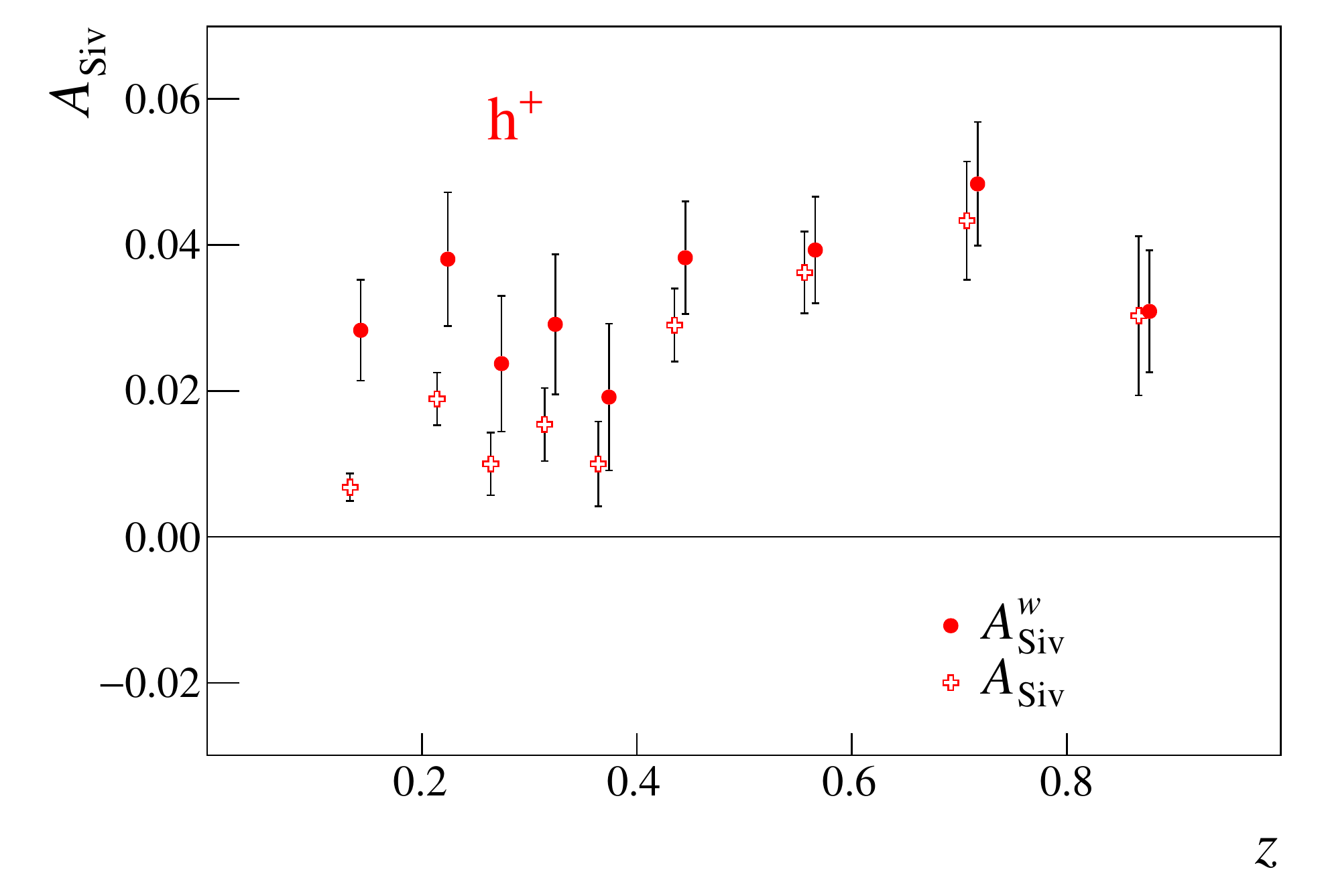}
\includegraphics[width=0.48\textwidth]{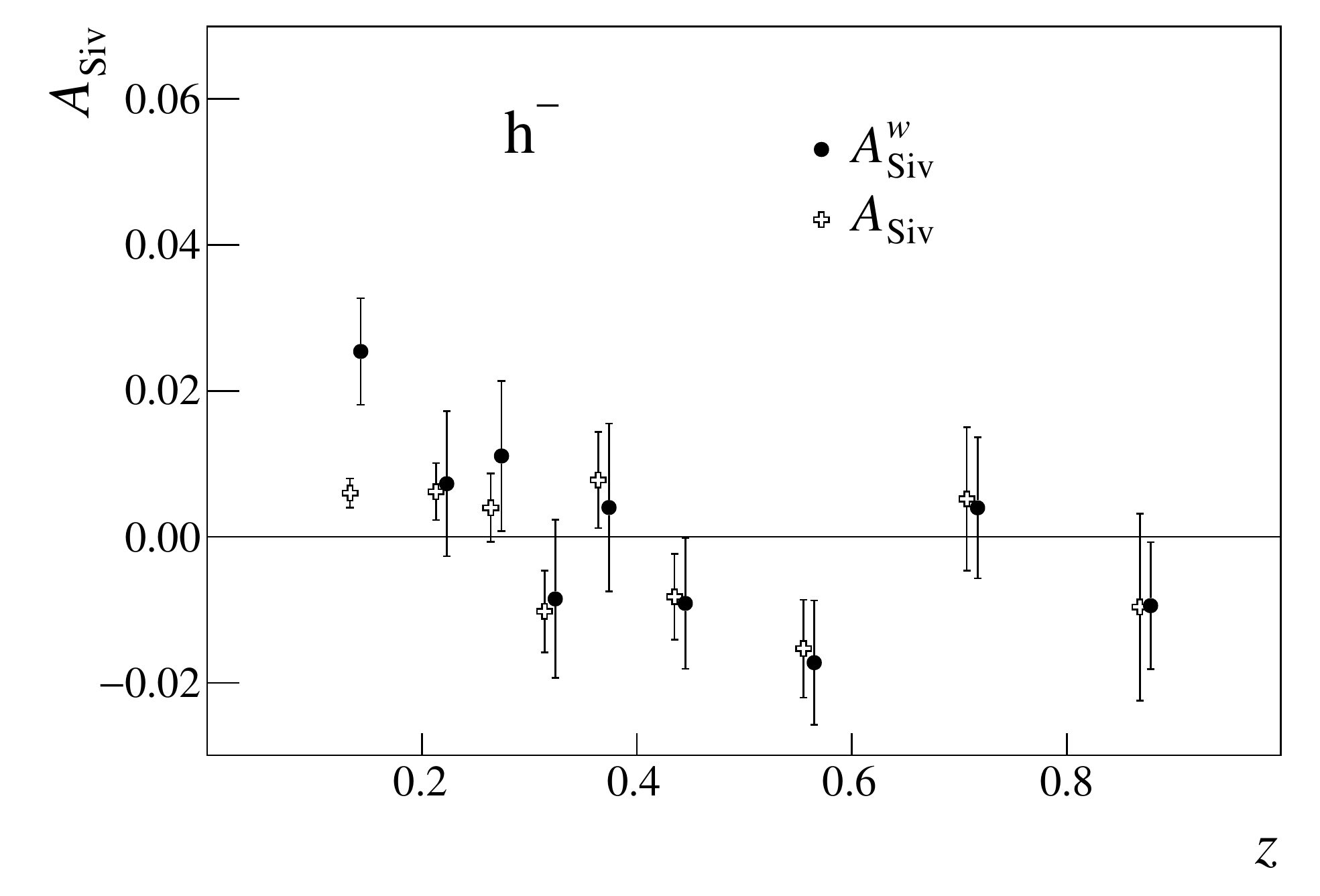}
\caption{Full points: $A_{\rm Siv}^{w}$ in the nine $z$ bins for positive (left
  panel) and negative (right panel) hadrons. The open crosses are the
  corresponding unweighted Sivers asymmetries $A_{\rm
    Siv}$~\cite{Adolph:2012sp}, which are slightly shifted towards smaller $x$ values for
  clarity.}
\label{fig:zbins}
\end{figure*}

In order to further investigate the $z$ dependence, it is of interest to look at
$A_{\rm Siv}^{w}$ as a function of $z$, after integration over $x$.
The results in the range $0.1<z<1$ are shown in Fig.~\ref{fig:zbins}.
For positive hadrons, the values are almost constant within statistical
uncertainties, as it is expected in the case of $u$-quark dominance if the
measurement is performed in the current-fragmentation region and factorisation
holds.
The values of the measured $P_T/zM$-weighted asymmetries are given in
Tables~\ref{tab:wa_xxx} and~\ref{tab:wa_xxxa}.

From the $P_T/zM$-weighted asymmetries it is straightforward to extract the
first transverse moment of the Sivers function.  This will be done in 
Section~\ref{sec:extraction}.

\section{Sivers asymmetries weighted by $P_T/M$}

\begin{figure*}[tb]
\centering
\includegraphics[width=0.48\textwidth]{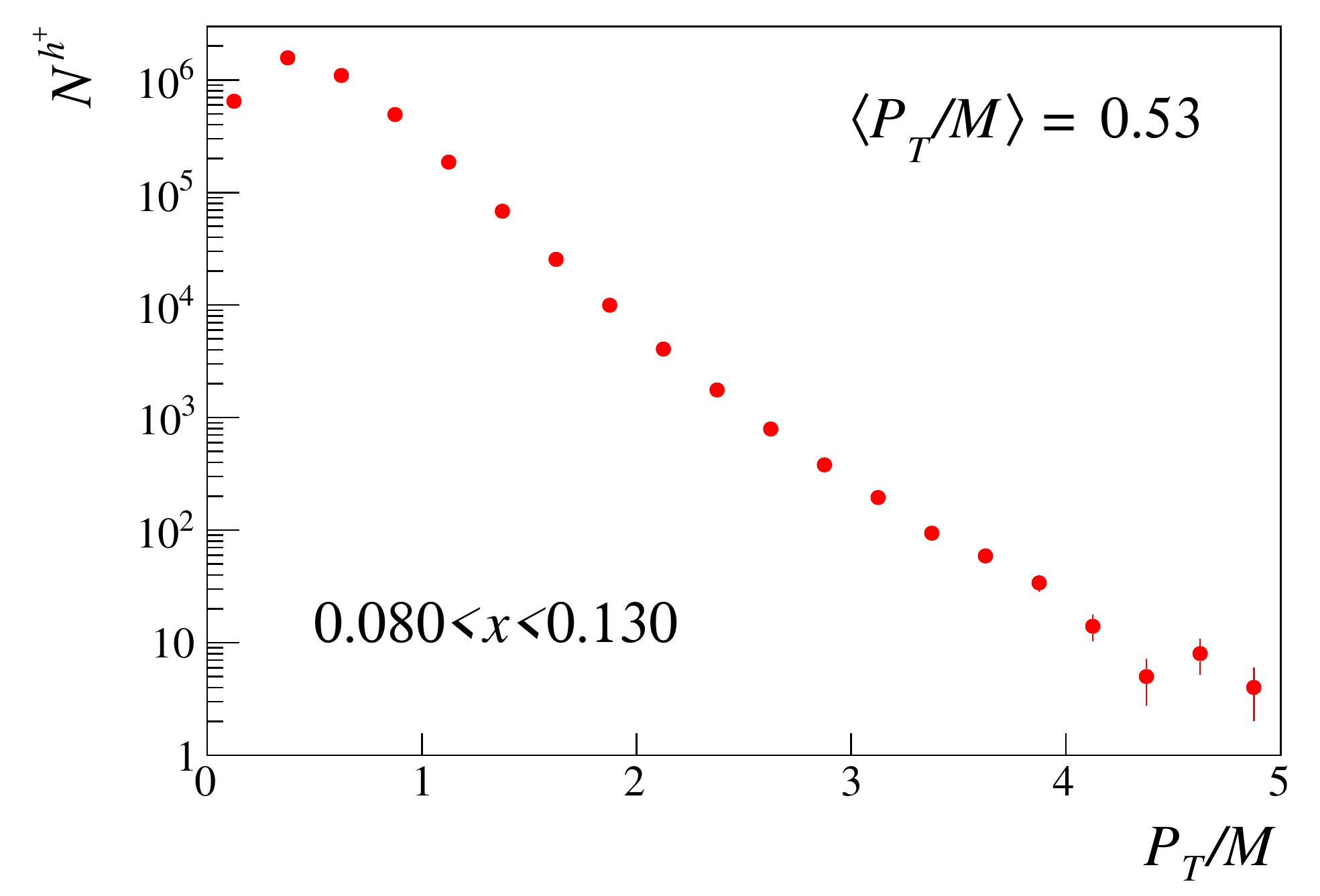}
\includegraphics[width=0.48\textwidth]{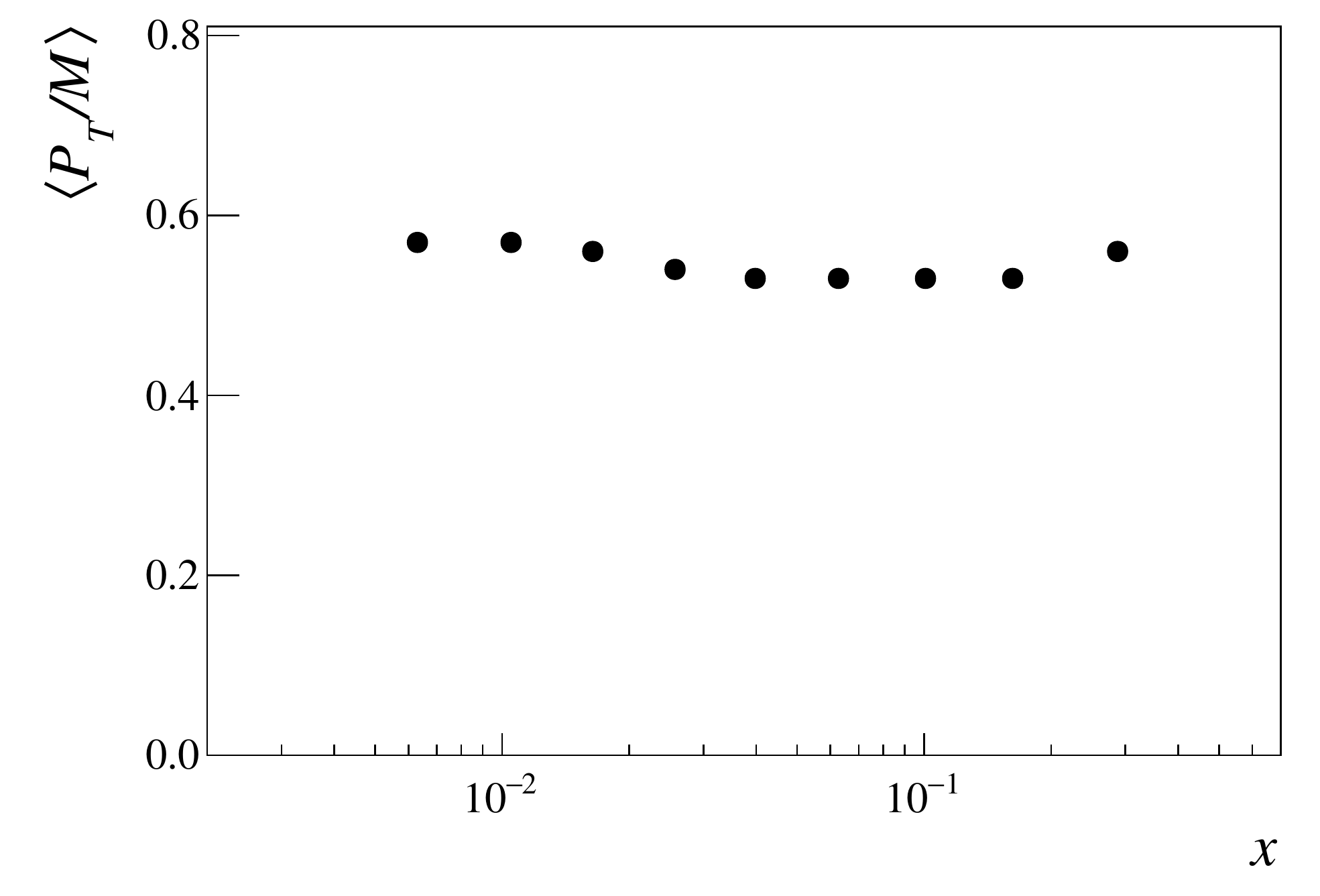}
\hfill
\caption{Left: distribution of the weight $w'=P_T/M$ for positive hadrons in the
  bin $0.080<x<0.13$.  Right: mean value of $w'$ as
  a function of $x$. No acceptance correction applied.}
\label{fig:weightspptz_x}
\end{figure*}

\begin{figure*}[tb]
\centering
\includegraphics[width=0.48\textwidth]{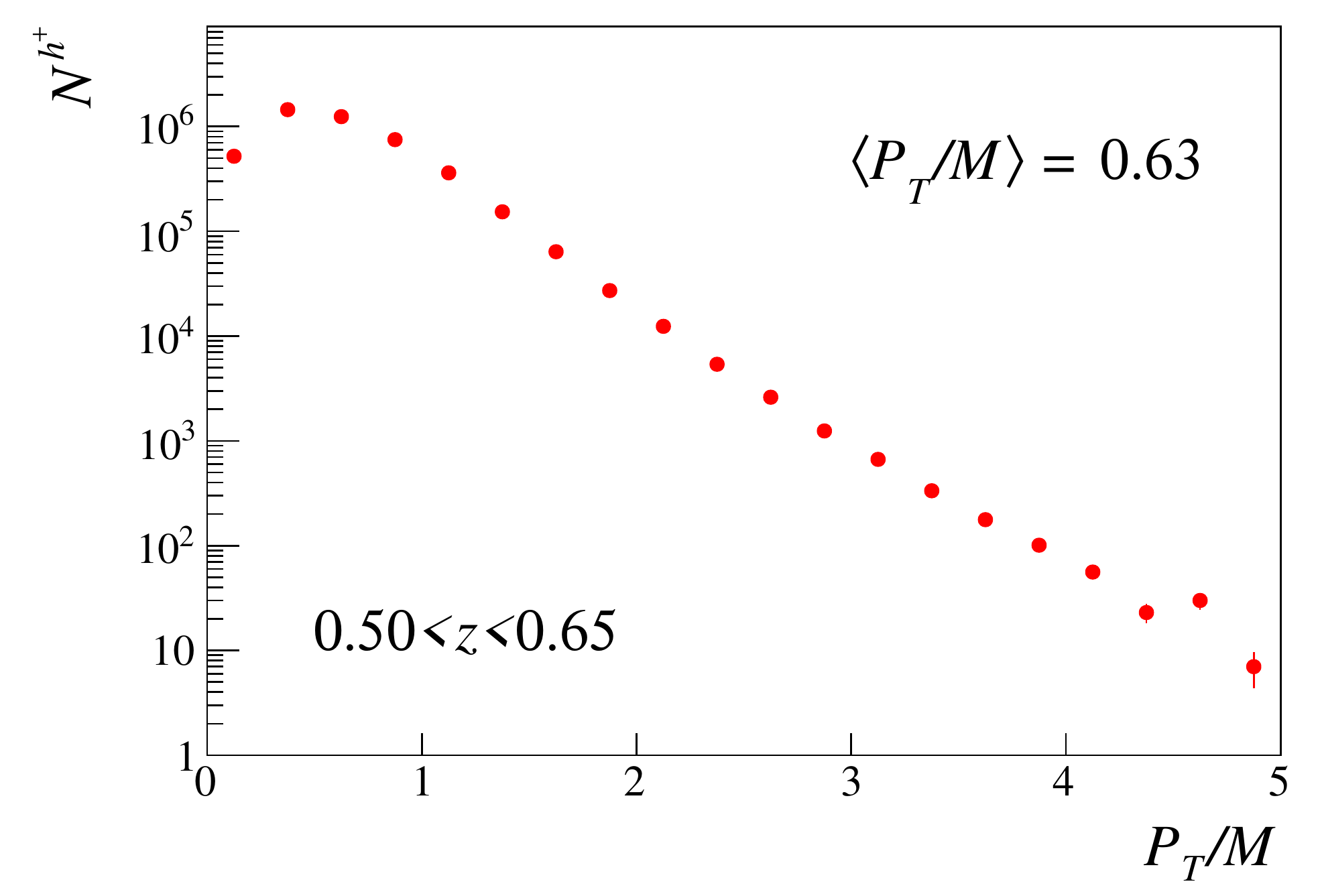}
\includegraphics[width=0.48\textwidth]{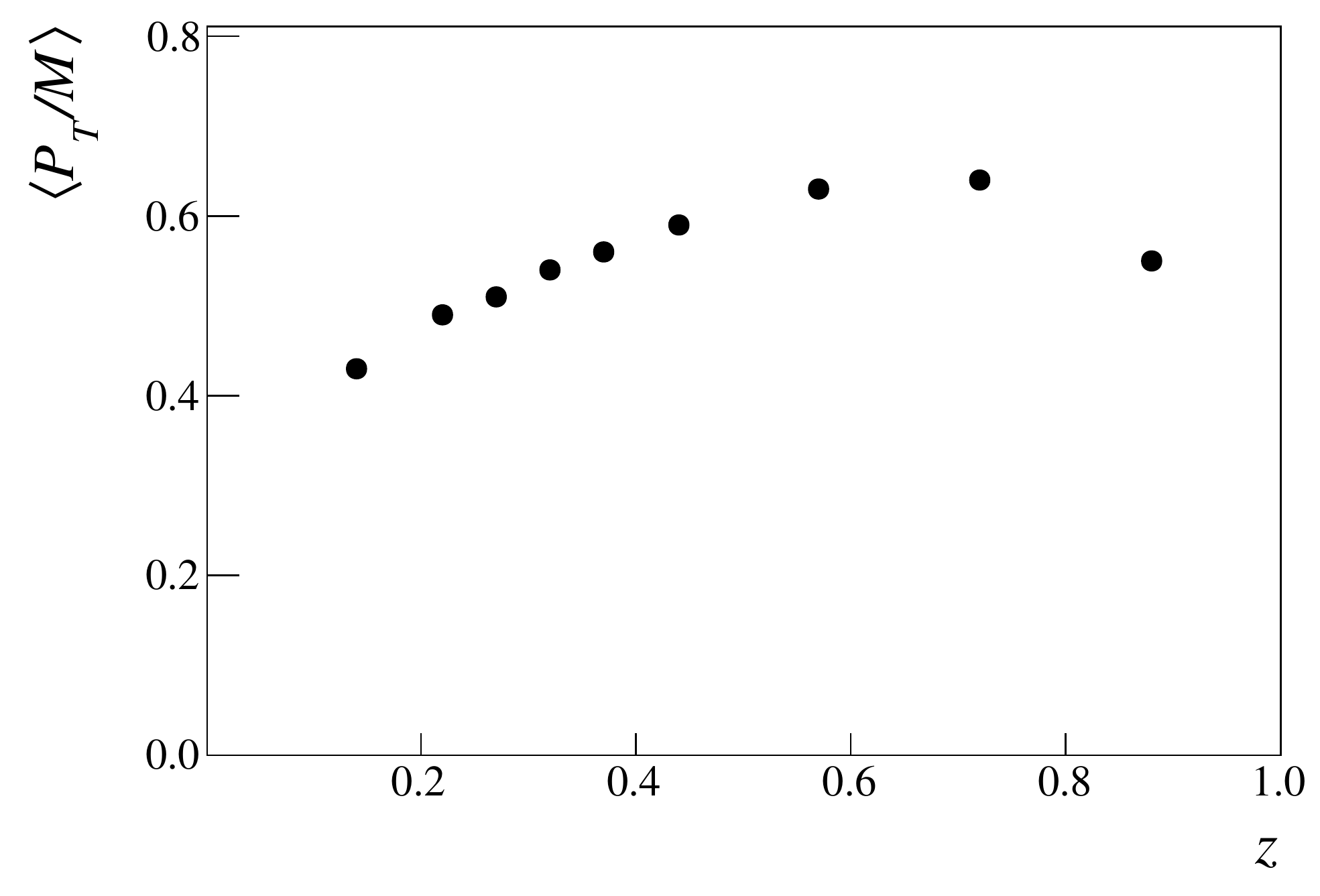}
\hfill
\caption{Left: Distribution of the weight $w'=P_T/M$ for positive hadrons in the
  bin $0.50<z<0.65$.  Right: mean value of $w'$ as
  a function of $z$.  No acceptance correction applied.}
\label{fig:weightspptz_z}
\end{figure*}

Let us now turn to the Sivers asymmetries weighted with $w'=P_T/M$.  The
distributions of $w'$ are very similar in all $x$ and $z$ bins.  Examples of the
distributions and the mean values of $w'$ in the $x$ and $z$ bins for positive hadrons are given in
Figs.~\ref{fig:weightspptz_x} and~\ref{fig:weightspptz_z}, respectively, and in
Table~\ref{tab:weightspptz}.
Again, for negative hadrons the distributions are very much the same.

\begin{figure}[tb]  
\centering
  \includegraphics[width=0.48\textwidth]{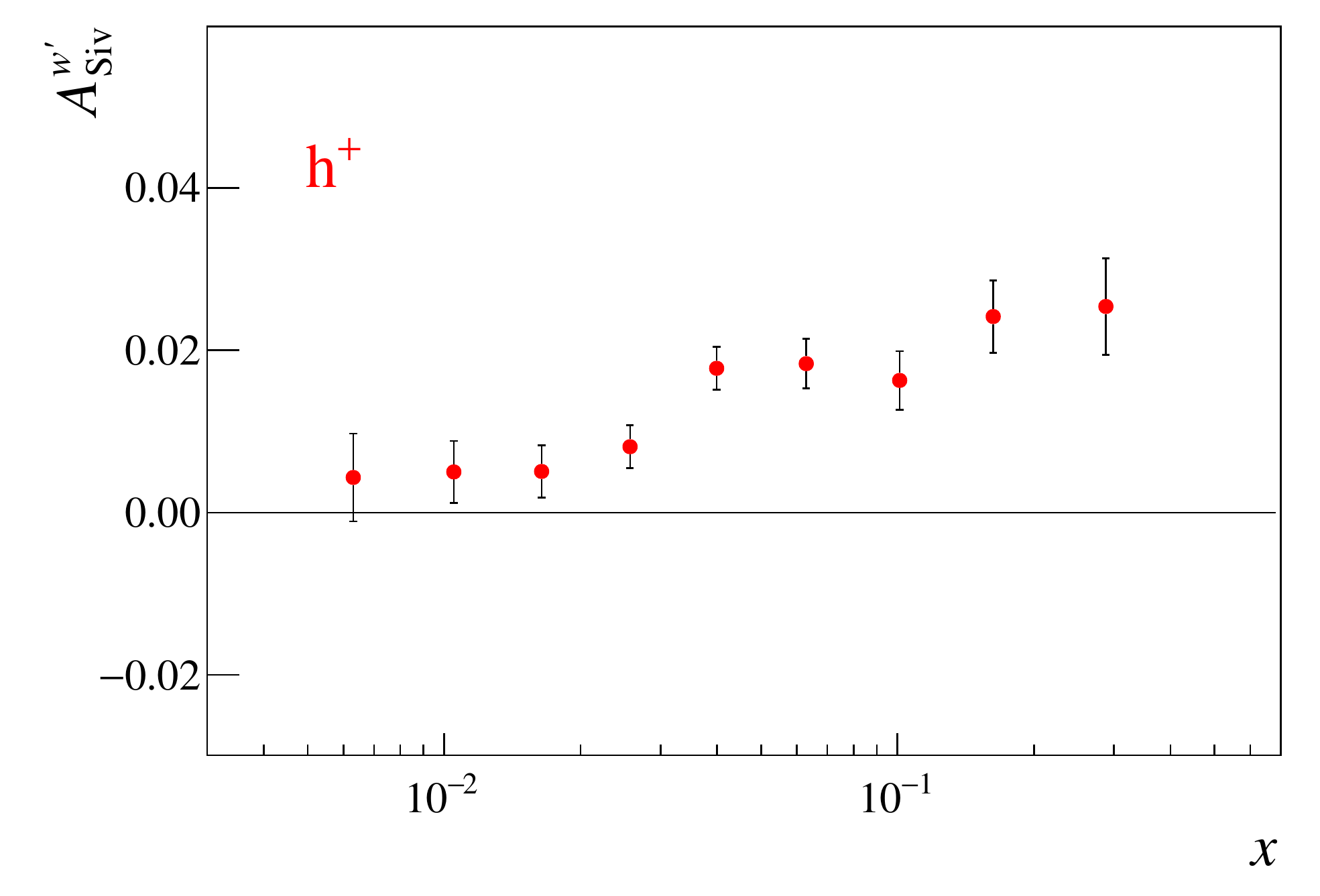}
  \includegraphics[width=0.48\textwidth]{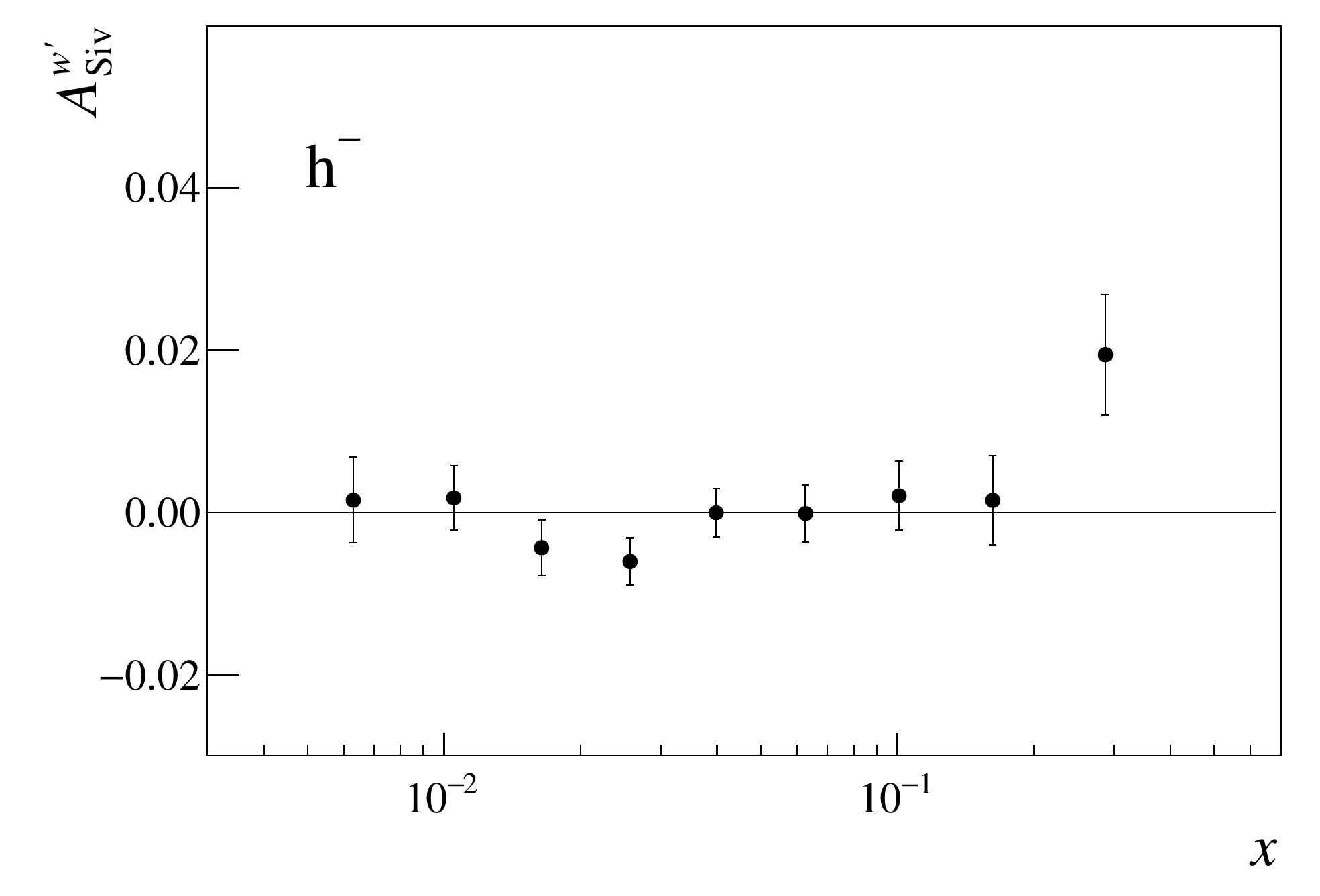}
\caption{ The weighted asymmetry $A^{w'}_{\rm Siv}$ with $w'=P_T^h/M$, as a function 
of $x$  for positive (left) and negative (right) 
hadrons with $z>0.2$.}
\label{fig:SDfig5} 
\end{figure}

\begin{figure}[tb]
\centering
\includegraphics[width=0.48\textwidth]{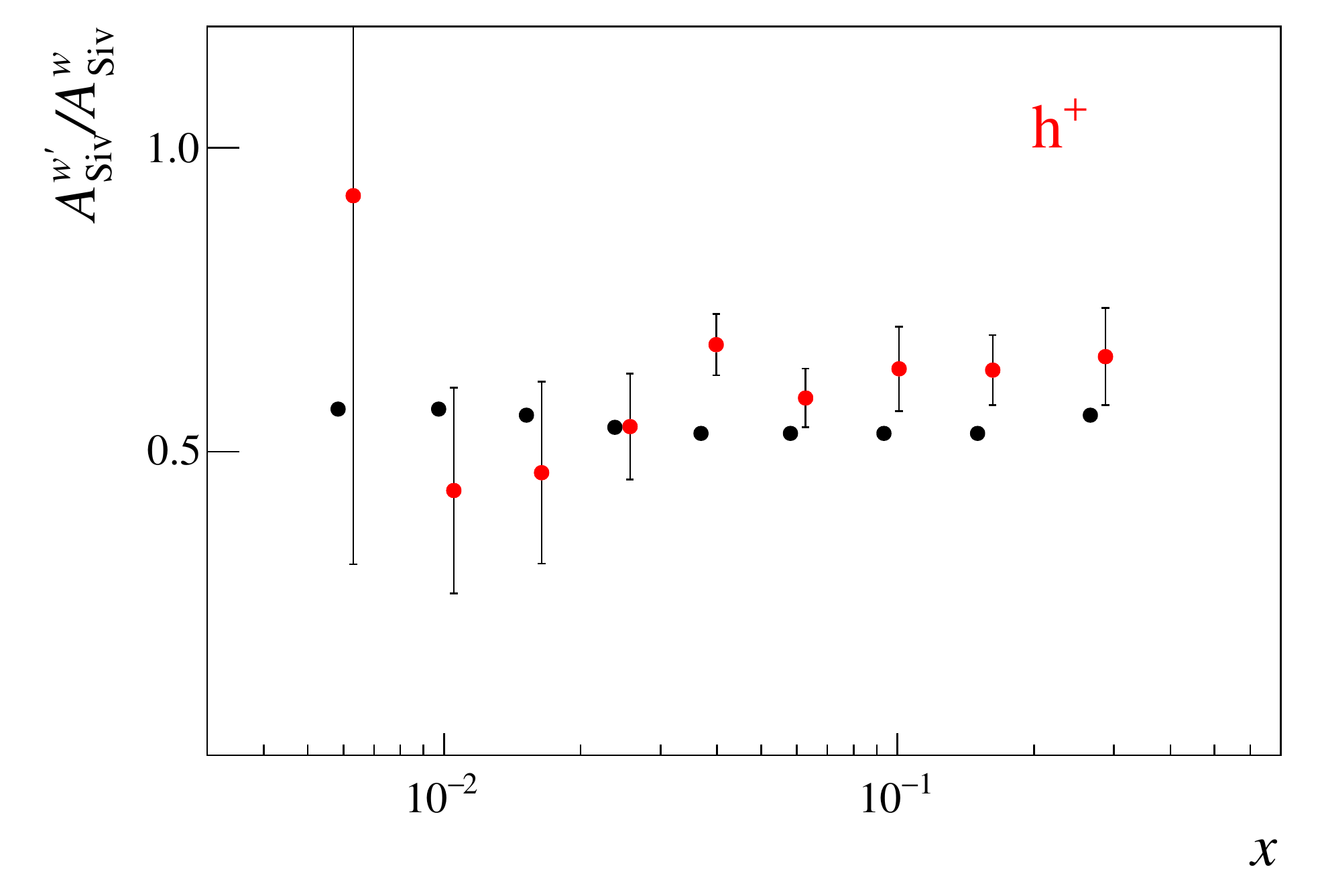}
\caption{Ratio $A^{w'}_{\rm Siv}/A_{\rm Siv}$ as a
function of $x$ for positive hadrons and $z>0.2$.
The black points are the values of 
$4 \langle  z \rangle / \pi M \langle  z/P_T \rangle $.}
\label{fig:SDfig7} 
\end{figure}

The results for $A^{w'}_{\rm Siv}$ are shown in Fig.~\ref{fig:SDfig5} for positive
and negative hadrons. 
The ratio $R^{w'}=A^{w'}_{\rm Siv}/A_{\rm Siv}$ for positive hadrons is shown in  
Fig.~\ref{fig:SDfig7}. 
Correlations between numerator and denominator were accounted for.
The ratio $R^{w'}=A^{w'}_{\rm Siv}/A_{\rm Siv}$ is almost constant as function of $x$ with a mean value of 0.62, not far from
that expected using the Gaussian model [see Eq.~(\ref{ratio})], which is also shown in the
figure.

\begin{figure}[tbp]%
\centering
\includegraphics[width=0.48\textwidth]{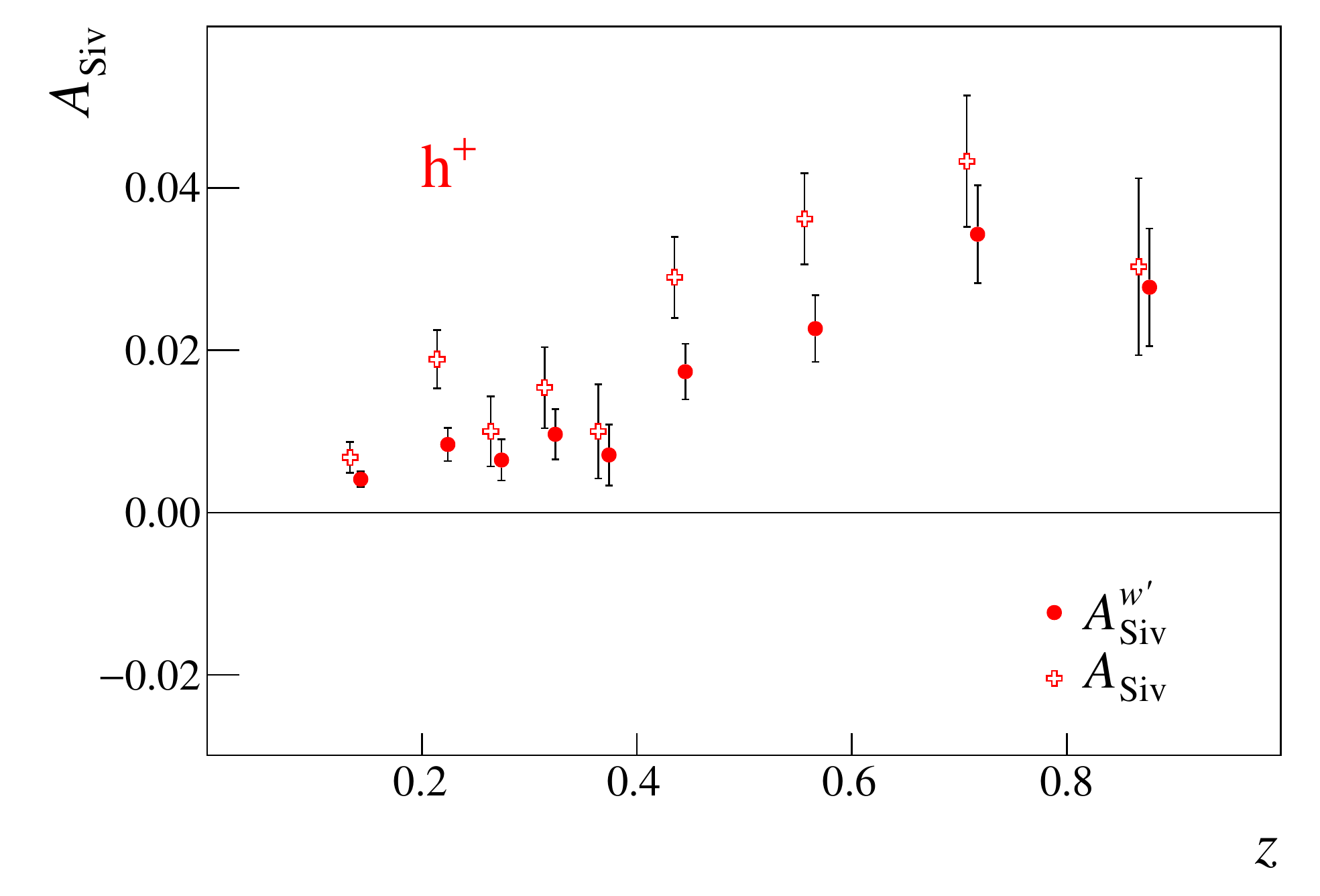}
\includegraphics[width=0.48\textwidth]{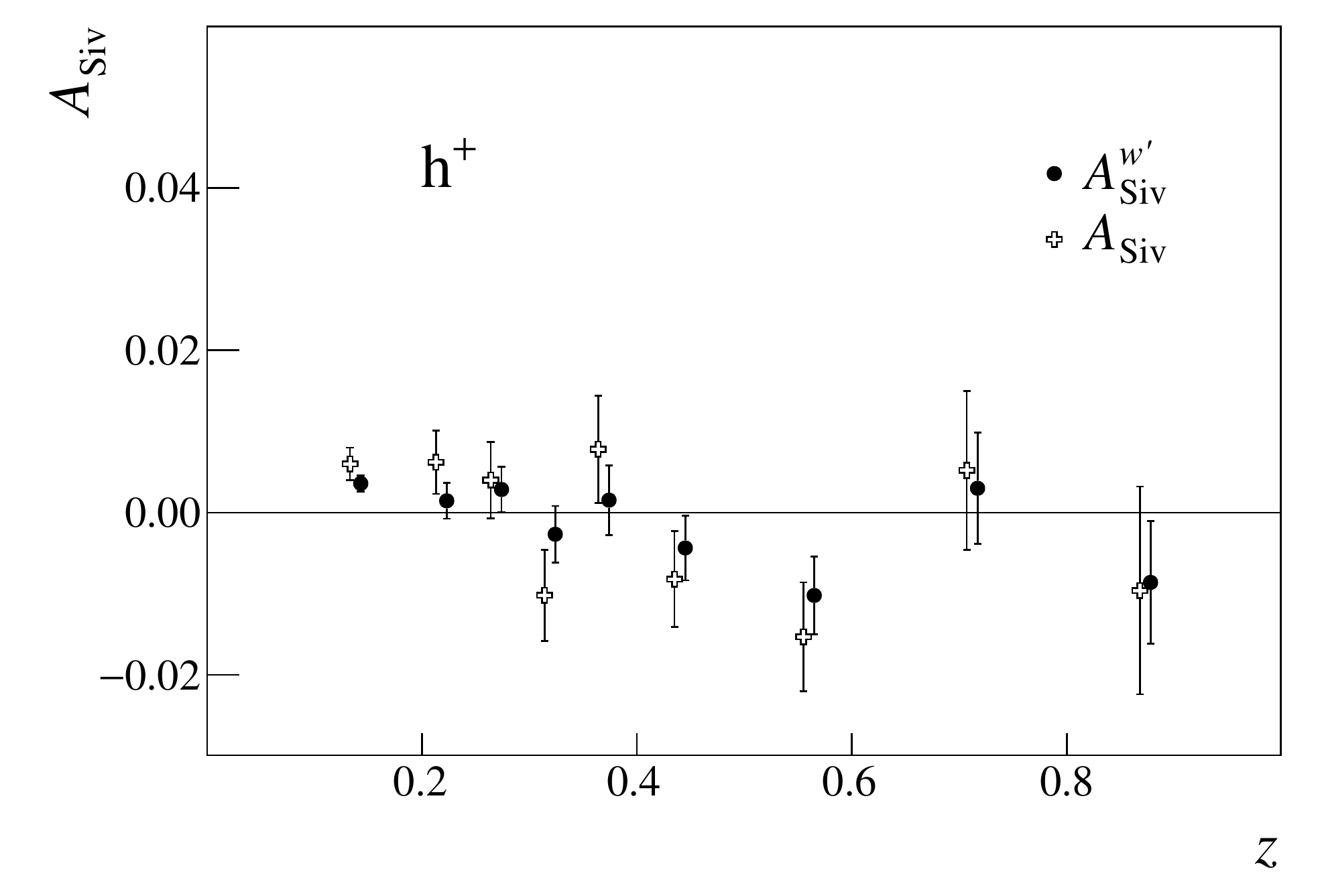}
\caption{Closed points: $A_{\rm Siv}^{w'}$ with $w'=P_T^h/M$ 
in the nine $x$ bins for positive (left panel) and negative (right panel)
hadrons. The open crosses are the unweighted Sivers asymmetries $A_{\rm
  Siv}$~\cite{Adolph:2012sp}, which are slightly shifted towards smaller $x$ values for
clarity.}
\label{fig:SDfig5_s2} 
\end{figure}

In order to better investigate the $z$ dependence, as in the case of the
$A^{w}_{\rm Siv}$ asymmetries, the analysis was repeated adding the hadrons
with $0.1 <z< 0.2$.  The results for the $x$-integrated asymmetry $A^{w'}_{\rm Siv}$
as a function of $z$ are shown in Fig.~\ref{fig:SDfig5_s2} for positive and
negative hadrons.
The values for positive hadrons are in qualitative agreement with the $u$-quark
dominance approximation, i.e.:
\be
A_{\rm Siv}^{w'}(z) \sim z .
\label{eq:SDwsa_zu}
\ee
For comparison, the published Sivers asymmetries $A_{\rm Siv}$~\cite{Adolph:2012sp}
are also shown in the same figure. All values of the measured $P_T/M$-weighted
asymmetries are given in Tables~\ref{tab:wa_xx} 
and~\ref{tab:wa_xxa}~\footnote{
All the numerical values for the results presented in this paper,
as well as the covariance matrices are available on HEPDATA~\cite{Durham}.
}
. 

\section{Point-by-point extraction of the first moments of the Sivers functions}
\label{sec:extraction}
The final goal of the measurement of the weighted Sivers asymmetries is the
extraction of the first moments of the Sivers functions.  
Thus we consider the
weighted asymmetry integrated over $z$ (we restore the $Q^2$ dependence):
\begin{eqnarray}
A_{\rm Siv}^{w}(x,Q^2) &=&
2 \, \frac{\sum_{q} e_q^2 x f_{1T}^{\perp \, (1) \, q}(x, Q^2) \, \widetilde{D}_1^q(Q^2)}
     {\sum_{q} e_q^2 x f_{1}^q(x, Q^2) \, \widetilde{D}_1^q(Q^2)} \,, 
\label{eq:integrated_wsa}
\end{eqnarray}
where 
\be
\widetilde{D}_{1}^q (Q^2) = 
\int_{z_{min}}^{z_{max}} \D z \, D_{1}^q (z, Q^2)\,.  
\ee
The denominator of Eq.~(\ref{eq:integrated_wsa}) can be fully evaluated by
resorting to global fits of distribution and fragmentation functions.

\begin{figure*}[tb]
\centering
\includegraphics[width=0.6\textwidth]{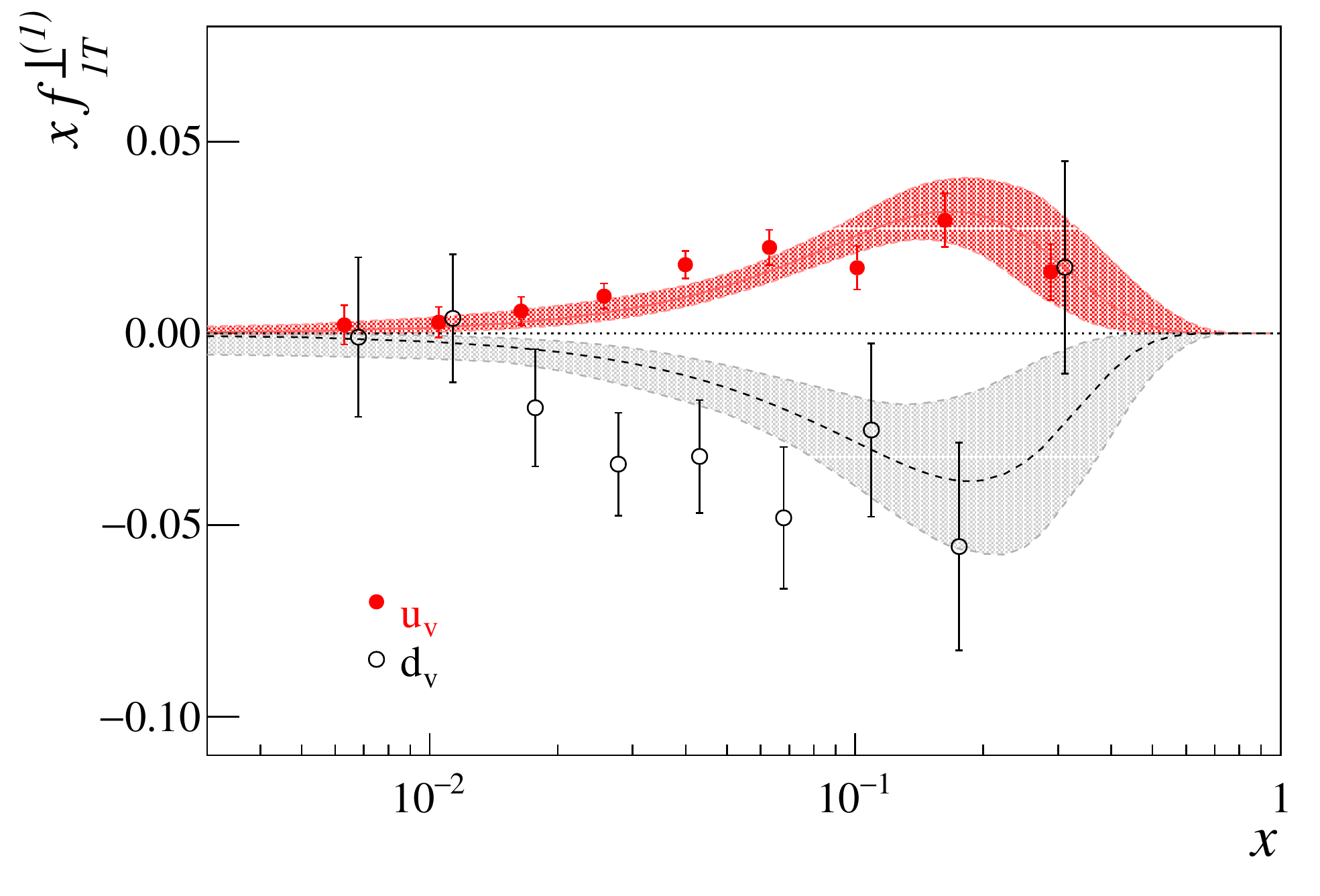}
\hfill
\caption{Values of the first moment of the Sivers function for  $u$ (closed
  red dots) and $d$ (open black dots) quarks from the $P_T/zM$ weighted-Sivers
  asymmetries for charged hadrons with $z>0.2$.  The curves and the uncertainty
  bands are the results of the fit of Ref.~\cite{Anselmino:2012aa}. }
\label{fig:uv_dv_cmp1}
\end{figure*}

There are two sets of asymmetries, i.e. for unidentified positively 
(superscript $+$)
and negatively (superscript $-$) charged hadrons. In our analysis, we omit 
the sea-quark Sivers distributions, which were shown to be
negligible in a previous study~\cite{Martin:2017yms}.  
The asymmetries then read (for simplicity we omit again
the $x$ and $Q^2$ dependence) 
\be
A_{\rm Siv}^{w, \pm} = 
2 \, \frac{4 x f_{1T}^{\perp \, (1) \, u_v} \, \widetilde{D}_{1}^{u, \pm}
+ x f_{1T}^{\perp \, (1) \, d_v} \, \widetilde{D}_{1}^{d, \pm}}{9 \, \sum_{q} 
e_q^2 x f_{1}^q \, \widetilde{D}_1^{q, \pm}} \,.  
\label{eq:explicit_wsa}
\ee
Denoting the denominator by $\delta^{\pm}$
\be
\delta^{\pm} \equiv 
9 \, \sum_{q} 
e_q^2 x f_{1}^q \, \widetilde{D}_1^{q, \pm}\,, 
\ee
the valence Sivers distributions can be extracted from the 
asymmetries as follows
\bq
x f_{1T}^{\perp \, (1) \, u_v} &=& \frac{1}{8} 
\, \frac{\delta^+ A_{\rm Siv}^{w, +} \widetilde{D}_{1}^{d,-}
- \delta^- A_{\rm Siv}^{w, -} \widetilde{D}_{1}^{d,+}}{ \widetilde{D}_{1}^{u,+}
\widetilde{D}_{1}^{d, -} - \widetilde{D}_{1}^{d, +}\widetilde{D}_{1}^{u, -}}
\,, 
\label{uval} \\
x f_{1T}^{\perp \, (1) \, d_v} &=& \frac{1}{2} 
\, \frac{\delta^- A_{\rm Siv}^{w, -} \widetilde{D}_{1}^{u,+}
- \delta^+ A_{\rm Siv}^{w, +} \widetilde{D}_{1}^{u,-}}{ \widetilde{D}_{1}^{u,+}
\widetilde{D}_{1}^{d, -} - \widetilde{D}_{1}^{d, +}\widetilde{D}_{1}^{u, -}}\,. 
\label{dval}
\eq
Eqs.~(\ref{uval}) and~(\ref{dval}) allow for a point-by-point extraction of the
Sivers distributions for valence quarks.  For the distribution functions we use the CTEQ5D
parametrisation~\cite{Brock:1993sz} and for the fragmentation functions of unidentified
hadrons the DSS parametrisation~\cite{deflorian:2007ekg}.  The results are
displayed in Fig.~\ref{fig:uv_dv_cmp1} and tabulated in Table~\ref{tab:valence}
together with the mean values of $Q^2$ (ranging from 1.24~$({\rm GeV}/c)^2$ to 
25.6~$({\rm GeV}/c)^2$).  
The extracted values for $x f_{1T}^{\perp \, (1) \, u_v}$ and $x
f_{1T}^{\perp \, (1) \, d_v}$ are correlated, as they are linear functions of the same
two measured asymmetries, and the computed correlation coefficients are also
given in Table~\ref{tab:valence}.

The uncertainties are computed from the statistical uncertainties 
of the measured
asymmetries, and no attempt was made to try to assign a systematic uncertainty
to the results.  The uncertainties in the extracted $d_v$ Sivers distribution
are much larger than the corresponding ones for the $u_v$ quark.  The $u_v$ and
$d_v$ Sivers distributions are linear combinations [see
  Eqs.~(\ref{uval},~\ref{dval})] of the same Sivers asymmetries for positive and
negative hadrons on the proton, thus in principle sufficient for their
determination, but the coefficient of proportionality is four times larger for
the $d$ quark, which makes the uncertainties of the extracted $x f_{1T}^{\perp
  \, (1) \, d_v}$ about four times larger than those of $x f_{1T}^{\perp \, (1)
  \, u_v}$.

In Fig.~\ref{fig:uv_dv_cmp1}, we also show for comparison the results, i.e. central
values and uncertainty bands, of the fit~\cite{Anselmino:2012aa} to the HERMES
proton data~\cite{Airapetian:2009ae} and the COMPASS proton and deuteron
data~\cite{Alekseev:2008aa,Adolph:2014zba}, which uses DGLAP evolution.  The
results are compatible, with a slightly different trend of $x f_{1T}^{\perp \,
  (1) \, d_v}$ suggested by the present extraction.

It is also interesting to compare our present result with the point-by-point
extraction of Ref.~\cite{Martin:2017yms}, where the pion Sivers asymmetries from
the COMPASS proton~\cite{Adolph:2014zba} and deuteron~\cite{Alekseev:2008aa}
data are used as input.  
The data set used in Ref.~\cite{Martin:2017yms} and the
present one have the dominating  pion data on the proton target
in common, so that the results are strongly correlated.
\begin{figure*}[tb]
\centering
\includegraphics[width=0.6\textwidth]{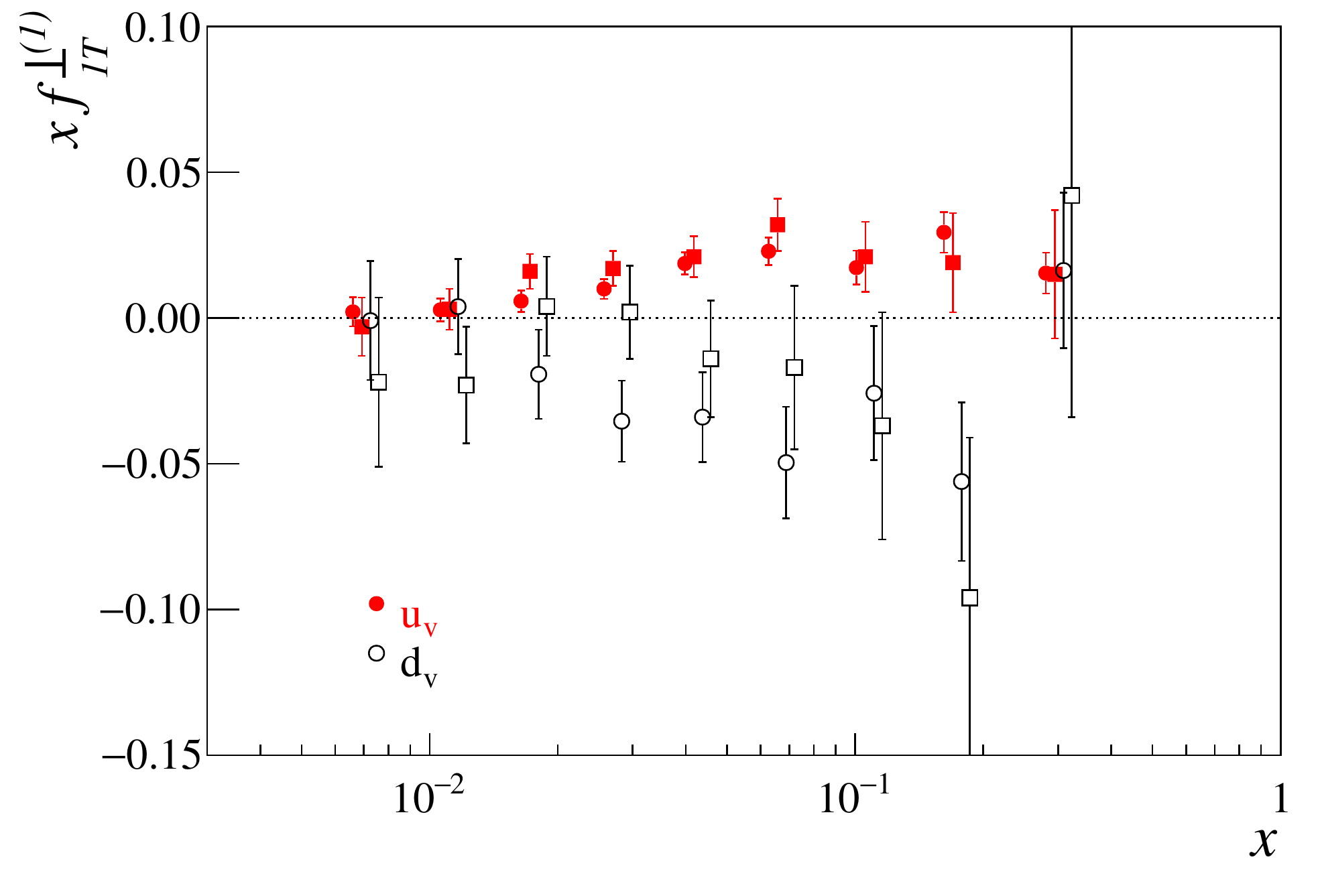}
\hfill
\caption{Comparison of the values of the first moment of the Sivers function for
  $u$ (closed red dots) and $d$ (open black dots) quarks from the
  $P_T/zM$-weighted Sivers asymmetries for charged hadrons with $z>0.2$, 
and the 
  corresponding values obtained in Ref.~\cite{Martin:2017yms} from the
  unweighted pion Sivers asymmetries measured by COMPASS on deuteron and proton
  (closed red and open black squares, respectively).}
\label{fig:uv_dv_cmp2}
\end{figure*}
As can be seen in Fig.~\ref{fig:uv_dv_cmp2}, in the present work 
the uncertainties on the extracted
$u_v$ and $d_v$ Sivers function moments are on average
smaller by a factor of about 1.5 with respect to the corresponding quantities in
Ref.~\cite{Martin:2017yms}.  This is due to the fact that in the present
analysis we had to assume the Sivers function of the sea quarks to be zero.
Following the method of Ref.~\cite{Martin:2017yms} and imposing the sea-quark
Sivers functions to be zero, we have determined the $u_v$ and $d_v$ functions
from the $\pi^+$ and $\pi^-$ 
proton asymmetries~\cite{Adolph:2014zba} only 
and verified that both the central values and the uncertainties are very
similar to the ones presented in this paper.  Thus the differences visible in
Fig.~\ref{fig:uv_dv_cmp2} can be attributed to the impact of the deuteron data
and to the extraction of the sea-quark Sivers function, rather then to the use
of unweighted asymmetries.
The assumption of a vanishing contribution from the sea quarks will be better
verified only when more neutron data will be available.

\section{Conclusions and outlook}
COMPASS has measured the weighted Sivers asymmetries 
in SIDIS of 160 GeV muons on
transversely polarised protons, extending the standard analysis of unweighted
asymmetries.  The weighted asymmetries were determined for positive and negative
hadrons using as weight either $P_T/zM$ or $P_T/M$.  In both cases, the
asymmetries were found to be positive for positive hadrons in the range
$x>0.013$ and compatible with zero for negative hadrons with $z>0.2$, very much
as in the case of the standard Sivers asymmetries.  
The $z$ dependence for positive
hadrons agrees with the expectation in the case of $u$-quark dominance and of a
measurement performed in the current-fragmentation region.

From the $P_T/zM$-weighted Sivers asymmetries, and under the hypothesis of
negligible Sivers functions for sea quarks, we have extracted the first moments
of the Sivers functions for $u_v$ and $d_v$ quarks. 
In the leading-order pQCD formalism, the obtained values are model
independent because of the use of  weighted asymmetries and because of the
point-by-point extraction. Previous model-dependent extractions that are
based on the Gaussian ansatz compare well with our results.

The present analysis hints at the validity of the Gaussian parametrisation for
the transverse-momentum dependence of the Sivers distribution function and the
fragmentation function, at least in the kinematic domain explored by our
measurement.  As in all other extractions of the Sivers functions from SIDIS
asymmetries on transversely polarized nucleons, the $d$-quark Sivers function
turns out to be poorly determined and strongly dependent on the assumptions on
the Sivers functions of the sea quarks.  This is due to the scarcity of Sivers
asymmetry data taken with a transversely polarised deuteron target, as compared
to the existing data taken with a transversely polarised proton target.  The
recently approved COMPASS run~\cite{addendum3} with a transversely polarised
deuteron target in 2021 is expected to allow for a much better extraction of the
Sivers functions for both quarks and antiquarks.

\clearpage

\begin{table}[b] 
\caption{Mean values of the weight $P_T/zM$ for positive hadrons in the
  nine bins of $x$ for $z>0.2$, and in the nine bins of $z$.}
\begin{center}
\begin{tabular}{|c|c|c|c|c|} 
\hline
 $\langle x \rangle$ & $\langle P_T/zM \rangle$ & & $\langle z \rangle$ & $\langle P_T/zM \rangle$ \\ 
\hline
0.0063 & 1.80 & &  0.14 &  3.13 \\
0.0105 & 1.77 & &  0.22 &  2.19 \\
0.0164 & 1.72 & &  0.27 &  1.88 \\
0.0257 & 1.65 & &  0.32 &  1.66 \\
0.0399 & 1.60 & &  0.37 &  1.50 \\
0.0629 & 1.59 & &  0.44 &  1.32 \\
0.101  & 1.58 & &  0.57 &  1.11 \\
0.163  & 1.56 & &  0.72 &  0.89 \\
0.288  & 1.62 & &  0.88 &  0.63 \\
\hline 
\end{tabular} 
\end{center}
\label{tab:weightsptz}
\end{table} 
\begin{table}[b] 
\caption{Measured values of the $P_T/zM$-weighted Sivers asymmetries 
in the nine $x$ bins.} 
\begin{center}
\begin{tabular}{|r|r|r|r|r|} 
\hline 
\multicolumn{1}{|c}{} & \multicolumn{2}{|c}{$z>0.2$} &
\multicolumn{2}{|c|}{$0.1<z<0.2$}\\
\hline 
\multicolumn{1}{|c}{$x$} & \multicolumn{1}{|c}{$A_{\rm Siv}^w, \, h^+$} 
&  \multicolumn{1}{|c}{$A_{\rm Siv}^w, \, h^-$} &  \multicolumn{1}{|c}{$A_{\rm Siv}^w, \, h^+$} & \multicolumn{1}{|c|}{$A_{\rm Siv}^w, \, h^-$} \\
\hline 
 0.003$-$0.008 & $ 0.009 \pm  0.018 $ & $  0.003 \pm  0.017 $  & $ -0.041 \pm  0.031 $ & $  0.001 \pm  0.032 $ \\ 
 0.008$-$0.013 & $ 0.013 \pm  0.012 $ & $  0.010 \pm  0.013 $  & $ -0.010 \pm  0.022 $ & $ -0.012 \pm  0.023 $ \\ 
 0.013$-$0.020 & $ 0.014 \pm  0.010 $ & $ -0.008 \pm  0.011 $  & $  0.030 \pm  0.019 $ & $  0.022 \pm  0.019 $ \\ 
 0.020$-$0.032 & $ 0.021 \pm  0.008 $ & $ -0.015 \pm  0.009 $  & $  0.050 \pm  0.015 $ & $  0.030 \pm  0.016 $ \\ 
 0.032$-$0.050 & $ 0.046 \pm  0.008 $ & $  0.004 \pm  0.010 $  & $  0.025 \pm  0.016 $ & $  0.044 \pm  0.017 $ \\ 
 0.050$-$0.080 & $ 0.050 \pm  0.009 $ & $ -0.002 \pm  0.011 $  & $  0.059 \pm  0.018 $ & $  0.016 \pm  0.019 $ \\ 
 0.080$-$0.130 & $ 0.039 \pm  0.011 $ & $  0.010 \pm  0.014 $  & $  0.044 \pm  0.021 $ & $  0.028 \pm  0.023 $ \\ 
 0.130$-$0.210 & $ 0.064 \pm  0.013 $ & $  0.007 \pm  0.017 $  & $  0.055 \pm  0.026 $ & $  0.092 \pm  0.029 $ \\ 
 0.210$-$0.700 & $ 0.063 \pm  0.017 $ & $  0.070 \pm  0.023 $  & $  0.012 \pm  0.034 $ & $  0.063 \pm  0.038 $ \\ 
\hline
\end{tabular} 
\end{center}
\label{tab:wa_xxx}
\end{table} 
\begin{table}[tb] 
\caption{Measured values of the $P_T/zM$-weighted Sivers asymmetries in the nine $z$ bins.} 
\begin{center}
\begin{tabular}{|r|r|r|r|r|} 
\hline 
\multicolumn{1}{|c}{$z$} & \multicolumn{1}{|c}{$A_{\rm Siv}^w, \, h^+$} & \multicolumn{1}{|c|}{$A_{\rm Siv}^w, \, h^-$} \\
\hline 
0.10$-$0.20 & $   0.0283 \pm    0.0069$& $   0.0254 \pm    0.0073$  \\ 
0.20$-$0.25 & $   0.0380 \pm    0.0091$& $   0.0073 \pm    0.0099$  \\ 
0.25$-$0.30 & $   0.0237 \pm    0.0093$& $   0.0111 \pm    0.0103$  \\ 
0.30$-$0.35 & $   0.0291 \pm    0.0096$& $  -0.0085 \pm    0.0108$  \\ 
0.35$-$0.40 & $   0.0191 \pm    0.0100$& $   0.0040 \pm    0.0115$  \\ 
0.40$-$0.50 & $   0.0382 \pm    0.0077$& $  -0.0091 \pm    0.0090$  \\ 
0.50$-$0.65 & $   0.0393 \pm    0.0073$& $  -0.0172 \pm    0.0085$  \\ 
0.65$-$0.80 & $   0.0483 \pm    0.0085$& $   0.0040 \pm    0.0097$  \\ 
0.80$-$1.00 & $   0.0309 \pm    0.0084$& $  -0.0094 \pm    0.0087$  \\ 
\hline
\end{tabular} 
\end{center}
\label{tab:wa_xxxa}
\end{table} 
\begin{table}[tb] 
\caption{Mean value of the weight $P_T/M$ for
positive hadrons in the nine $x$ bins 
 for $z>0.2$,  and in the nine $z$ bins.} 
\begin{center}
\begin{tabular}{|c|c|c|c|c|c|} 
\hline
 $\langle x \rangle$ & $\langle P_T/M \rangle$ & & $\langle z \rangle$ & $\langle P_T/M \rangle$ \\ 
\hline
0.0063 & 0.57 & &   0.14 & 0.43  \\
0.0105 & 0.57 & &   0.22 & 0.49  \\
0.0164 & 0.56 & &   0.27 & 0.51  \\
0.0257 & 0.54 & &   0.32 & 0.54  \\
0.0399 & 0.53 & &   0.37 & 0.56  \\
0.0629 & 0.53 & &   0.44 & 0.59  \\
0.101  & 0.53 & &   0.57 & 0.63  \\
0.163  & 0.53 & &   0.72 & 0.64  \\
0.288  & 0.56 & &   0.88 & 0.55  \\
\hline 
\end{tabular} 
\end{center}
\label{tab:weightspptz}
\end{table} 
\begin{table}[tb] 
\caption{Measured values of the $P_T/M$-weighted Sivers asymmetries in the nine $x$ bins.} 
\begin{center}
\begin{tabular}{|r|r|r|r|r|} 
\hline 
\multicolumn{1}{|c}{} & \multicolumn{2}{|c}{$z>0.2$} &
\multicolumn{2}{|c|}{$0.1<z<0.2$}\\
\hline 
\multicolumn{1}{|c}{$x$} & \multicolumn{1}{|c}{$A_{\rm Siv}^{w'}, \, h^+$} 
&  \multicolumn{1}{|c}{$A_{\rm Siv}^{w'}, \, h^-$} &  \multicolumn{1}{|c}{$A_{\rm Siv}^{w'}, \, h^+$} & \multicolumn{1}{|c|}{$A_{\rm Siv}^{w'}, \, h^-$} \\
\hline 
 0.003$-$0.008 & $  0.0094  \pm    0.0175$ & $   0.0034  \pm    0.0172$ & $  -0.0407  \pm    0.0314$   & $  0.0007  \pm    0.0318$  \\ 
 0.008$-$0.013 & $  0.0135  \pm    0.0123$ & $   0.0100  \pm    0.0129$ & $  -0.0096  \pm    0.0221$   & $ -0.0117  \pm    0.0227$  \\ 
 0.013$-$0.020 & $  0.0143  \pm    0.0103$ & $  -0.0083  \pm    0.0112$ & $   0.0297  \pm    0.0187$   & $  0.0217  \pm    0.0194$  \\ 
 0.020$-$0.032 & $  0.0212  \pm    0.0083$ & $  -0.0147  \pm    0.0093$ & $   0.0496  \pm    0.0155$   & $  0.0299  \pm    0.0162$  \\ 
 0.032$-$0.050 & $  0.0462  \pm    0.0083$ & $   0.0037  \pm    0.0095$ & $   0.0246  \pm    0.0157$   & $  0.0437  \pm    0.0168$  \\ 
 0.050$-$0.080 & $  0.0495  \pm    0.0094$ & $  -0.0016  \pm    0.0113$ & $   0.0591  \pm    0.0180$   & $  0.0164  \pm    0.0194$  \\ 
 0.080$-$0.130 & $  0.0393  \pm    0.0110$ & $   0.0100  \pm    0.0137$ & $   0.0436  \pm    0.0212$   & $  0.0284  \pm    0.0230$  \\ 
 0.130$-$0.210 & $  0.0640  \pm    0.0135$ & $   0.0074  \pm    0.0175$ & $   0.0551  \pm    0.0264$   & $  0.0924  \pm    0.0291$  \\ 
 0.210$-$0.700 & $  0.0630  \pm    0.0174$ & $   0.0701  \pm    0.0234$ & $   0.0115  \pm    0.0343$   & $  0.0634  \pm    0.0382$  \\ 
\hline
\end{tabular} 
\end{center}
\label{tab:wa_xx}
\end{table} 
\begin{table}[tb] 
\caption{Measured values of the $P_T/M$-weighted Sivers asymmetries in the nine $z$ bins.} 
\begin{center}
\begin{tabular}{|r|r|r|r|r|} 
\hline 
\multicolumn{1}{|c}{$z$} & \multicolumn{1}{|c|}{$A_{\rm Siv}^{w'}, \, h^+$} & \multicolumn{1}{c|}{$A_{\rm Siv}^{w'}, \, h^-$} \\
\hline 
0.01$-$0.20 & $0.0041 \pm 0.0010$& $   0.0036 \pm    0.0010$  \\ 
0.20$-$0.25 & $0.0084 \pm 0.0020$& $   0.0015 \pm    0.0022$  \\ 
0.25$-$0.30 & $0.0065 \pm 0.0025$& $   0.0028 \pm    0.0028$  \\ 
0.30$-$0.35 & $0.0097 \pm 0.0031$& $  -0.0027 \pm    0.0035$  \\ 
0.35$-$0.40 & $0.0071 \pm 0.0037$& $   0.0015 \pm    0.0043$  \\ 
0.40$-$0.50 & $0.0173 \pm 0.0034$& $  -0.0044 \pm    0.0040$  \\ 
0.50$-$0.65 & $0.0227 \pm 0.0041$& $  -0.0102 \pm    0.0048$  \\ 
0.65$-$0.80 & $0.0343 \pm 0.0060$& $   0.0030 \pm    0.0069$  \\ 
0.80$-$1.00 & $0.0278 \pm 0.0072$& $  -0.0086 \pm    0.0075$  \\ 
\hline
\end{tabular} 
\end{center}
\label{tab:wa_xxa}
\end{table} 
\begin{table}[tb] 
\caption{Values of the first moments of the Sivers functions
for $u$ and $d$ quarks.
The last column gives their correlation coefficient $\rho$.} 
\begin{center}
\begin{tabular}{|r|r|r|r|r|} 
\hline
 \multicolumn{1}{|c}{$\langle x \rangle$} & \multicolumn{1}{|c}{$\langle Q^2 \rangle$} & \multicolumn{1}{|c}{$x f_{1T}^{\perp \, (1) \, u_v}$} &  \multicolumn{1}{|c}{$x f_{1T}^{\perp \, (1) \, d_v}$} 
& \multicolumn{1}{|c|}{$\rho$} \\ 
  & $({\rm GeV}/c)^2$ &  &  
&  \\ 
\hline 
0.0063 &   1.27 & $0.0022 \pm 0.0051$ & $-0.001 \pm  0.021$ &  $-0.26$ \\
0.0105 &   1.55 & $0.0029 \pm 0.0040$ & $ 0.004 \pm  0.017$ &  $-0.31$ \\
0.0164 &   1.83 & $0.0058 \pm 0.0037$ & $-0.019 \pm  0.015$ &  $-0.37$ \\
0.0257 &   2.17 & $0.0097 \pm 0.0033$ & $-0.034 \pm  0.013$ &  $-0.43$ \\
0.0399 &   2.82 & $0.0179 \pm 0.0036$ & $-0.032 \pm  0.015$ &  $-0.52$ \\
0.0629 &   4.34 & $0.0224 \pm 0.0046$ & $-0.048 \pm  0.019$ &  $-0.63$ \\
0.101  &   6.76 & $0.0171 \pm 0.0057$ & $-0.025 \pm  0.023$ &  $-0.68$ \\
0.163  &  10.6 & $0.0295 \pm 0.0070$ & $-0.056 \pm  0.027$ &  $-0.65$ \\
0.288  &  20.7 & $0.0160 \pm 0.0073$ & $ 0.017 \pm  0.028$ &  $-0.40$ \\
\hline
\end{tabular} 
\end{center}
\label{tab:valence}
\end{table} 

\clearpage

\section*{Appendix A - 
Transverse-momentum convolution in the weighted asymmetry}

The Sivers asymmetry weighted with the factor $w = P_T/zM$ reads  
\be
 A_{\rm Siv}^w(x,z) 
 = 
\frac{\sum_{q} e_q^2 x \, \int \D^2 \Vec P_T 
\, \frac{P_T}{z M} \, \mathcal{C} \left [ 
\frac{\Vec P_T \cdot \Vec k_T}{M P_T} \,   
f_{1T}^{\perp  \, q}(x, k_T^2) \, D_1^q(z, p_T^2) \right ] }
     {\sum_{q} e_q^2 x \,  f_{1}^q(x) D_1^q(z) } .  
\label{eq:w_sa_appendix}
\ee
The numerator contains the integral 
\bq
& & \int \D^2 \Vec P_T \, 
\frac{P_T}{z M} \,  \mathcal{C}  \left [  \frac{\Vec P_T \cdot \Vec k_T}{M P_T} \,   
f_{1T}^{\perp  \, q} \, D_1^q  \right ] 
\nonumber \\
& & \;\;\;\;
= 
\int \D^2 \Vec P_T \, \frac{P_T}{z M} 
\, \int \D^2 \Vec k_T  \int \D^2 \Vec p_T \, 
\delta^2 (z \Vec k_T + \Vec p_T - \Vec P_T) \, 
\frac{\Vec P_T \cdot \Vec k_T}{M P_T} \,  
\nonumber \\
& & \;\;\;\;\;\;\;\;\;\; 
f_{1T}^{\perp  \, q}(x, k_T^2) \, D_1^q(z, p_T^2)\,.  
\label{conv_app}
\eq
Using the delta function to integrate over $\Vec P_T$ gives
\bq
& & \int \D^2 \Vec P_T \, 
\frac{P_T}{z M} \,  \mathcal{C}  \left [  \frac{\Vec P_T \cdot \Vec k_T}{M P_T} \,   
f_{1T}^{\perp  \, q} \, D_1^q  \right ] 
\nonumber \\
& & \;\;\;\;
=  
 \int \D^2 \Vec k_T  \, \frac{1}{z M^2} \,  \int \D^2 \Vec p_T \,  
(z k_T^2 + \Vec k_T \cdot \Vec p_T) \, 
f_{1T}^{\perp  \, q}(x, k_T^2) \, D_1^q(z, p_T^2)\,
\nonumber \\
& & \;\;\;\; =
\int \D^2 \Vec k_T  \, \frac{k_T^2}{M^2} \, f_{1T}^{\perp  \, q}(x, k_T^2) 
\, \int \D^2 \Vec p_T \,  
  D_1^q(z, p_T^2)\,
\nonumber \\
& & \;\;\;\;=
2 \, f_{1T}^{\perp (1) q} (x) \, D_1^{q}(z) \,. 
\label{conv_app_2}
\eq

\bibliographystyle{npb_short}
\bibliography{articlebib}

\end{document}